\documentclass[prb,twocolumn,superscriptaddress,citeautoscript,showpacs,amsart]{revtex4-1}

\usepackage{graphicx}
\usepackage{multirow}
\usepackage{color}
\usepackage{bm}
\usepackage{times}
\usepackage{amsmath,bm,amsfonts}
\usepackage{dcolumn}
\usepackage{graphicx}
\usepackage{latexsym}
\usepackage{cancel}
\usepackage{hyperref}

\usepackage{ulem} 
\usepackage{braket}
\usepackage{xpatch}

\makeatletter
\patchcmd{\@ssect@ltx}
{\addcontentsline{toc}{#1}{\protect\numberline{}#8}}
{}
{}
{}
\makeatother

\def\n{\nonumber \\ }

\newcommand{\red}[1]{{\textcolor{red}{#1}}}

\newcommand{\ma}{\sigma}
\newcommand{\V}{\vec}
\newcommand{\dt}{\delta}
\newcommand{\ep}{\epsilon}
\newcommand{\f}{\frac}
\newcommand{\ta}{\theta}
\newcommand{\Om}{\Omega}
\newcommand{\pa}{\partial}
\newcommand{\Dt}{\Delta}
\newcommand{\da}{\dagger}
\newcommand{\kb}{\mathbf{k}}
\newcommand{\Mb}{\mathbf{M}}

\newcommand{\Db}{\mathbf{D}}
\newcommand{\Sb}{\mathbf{S}}
\newcommand{\Bb}{\mathbf{B}}
\newcommand{\Eb}{\mathbf{E}}

\newcommand{\vb}{\mathbf{v}}
\newcommand{\Omb}{\mathbf{\Omega}}
\renewcommand{\thefigure}{\arabic{figure}}
\renewcommand{\thetable}{\arabic{table}}

\renewcommand{\arraystretch}{1.5}
\newcommand{\A}{\alpha}
\newcommand{\B}{\beta}

\begin{document}

\title{Theory of transverse magnetization in spin-orbit coupled antiferromagnets}

\author{Taekoo \surname{Oh}}
\affiliation{Department of Physics and Astronomy, Seoul National University, Seoul 08826, Korea}

\affiliation{Center for Correlated Electron Systems, Institute for Basic Science (IBS), Seoul 08826, Korea}

\affiliation{Center for Theoretical Physics (CTP), Seoul National University, Seoul 08826, Korea}

\author{Sungjoon \surname{Park}}
\affiliation{Department of Physics and Astronomy, Seoul National University, Seoul 08826, Korea}

\affiliation{Center for Correlated Electron Systems, Institute for Basic Science (IBS), Seoul 08826, Korea}

\affiliation{Center for Theoretical Physics (CTP), Seoul National University, Seoul 08826, Korea}

\author{Bohm-Jung \surname{Yang}}
\email{bjyang@snu.ac.kr}
\affiliation{Department of Physics and Astronomy, Seoul National University, Seoul 08826, Korea}

\affiliation{Center for Correlated Electron Systems, Institute for Basic Science (IBS), Seoul 08826, Korea}

\affiliation{Center for Theoretical Physics (CTP), Seoul National University, Seoul 08826, Korea}

\date{\today}

\begin{abstract}
Some antiferromagnets under a magnetic field develop magnetization perpendicular to the field as well as more conventional ones parallel to the field.
So far, the transverse magnetization (TM) has been attributed to either spin canting effect or the presence of cluster magnetic multipolar ordering.
However, a general theory of TM based on microscopic understanding is still missing. 
Here, we construct a general microscopic theory of TM in antiferromagnets with cluster magnetic multipolar ordering by considering classical spin Hamiltonians with spin anisotropy that arises from the spin-orbit coupling. 
First, from general symmetry analysis, we show that TM can appear only when all crystalline symmetries are broken other than the antiunitary mirror, antiunitary two-fold rotation, and inversion symmetries. 
Moreover, by analyzing spin Hamiltonians, we show that TM always appears when the degenerate ground state manifold of the spin Hamiltonian is discrete. 
On the other hand, when the degenerate ground state manifold is continuous, TM generally does not appear except when the magnetic field direction and the spin configuration satisfy specific geometric conditions under single-ion anisotropy. 
Finally, we show that TM can induce anomalous planar Hall Effect, a unique transport phenomenon that can be used to probe multipolar antiferromagnetic structures.
We believe that our theory provides a useful guideline for understanding the anomalous magnetic responses of the antiferromagnets with complex magnetic structures.
\end{abstract}


\maketitle

{\it Introduction.|}
Spin-orbit coupled antiferromagnets are a promising playground to study novel correlated topological states and anomalous transport phenomena~\cite{tokura2019magnetic,witczak2014correlated}.
The complex spin structures of spin-orbit coupled antiferromagnets
can be characterized by their cluster magnetic multipole (CMM) moments reflecting the symmetry of the magnetic ground state~\cite{suzuki2017cluster,suzuki2019multipole}. Especially, those with higher-rank CMMs can exhibit anomalous transport phenomena including various types of anomalous Hall effects~\cite{suzuki2017cluster,suzuki2019multipole, vsmejkal2020crystal, gao2018orbital, matsumoto2014thermal,mishchenko2014equilibrium,zyuzin2016magnon,cheng2016spin,park2020thermal,kim2018unconventional,ueda2017magnetic,ueda2018spontaneous,ohtsuki2019strain,zhang2018spin}. The distinct magnetic symmetry of higher-rank CMMs underlies their unconventional physical properties, unexpected in simple spin systems with magnetic dipoles only.

Normally, when a magnetic field $\Bb$ is applied to an antiferromagnet, the magnetization is developed along the field direction. However, in several antiferromagnets with spin anisotropy including Gd$_2$Ti$_2$O$_7$, CsMnBr$_3$, and Eu$_2$Ir$_2$O$_7$,~\cite{abarzhi1992spin,glazkov2005single,glazkov2006observation,glazkov2007single,liang2017orthogonal,li2021correlated}, 
transverse magnetization (TM) was also observed. 
More specifically, in Gd$_2$Ti$_2$O$_7$ and CsMnBr$_3$, TM was observed when $\Bb$ was along certain directions and
was attributed to the spin canting effect.
More recently, TM was also observed in Eu$_2$Ir$_2$O$_7$. But in this system,
the presence of a magnetic octupolar ordering, not the spin canting effect,
was proposed as the origin of TM based on phenomenological Landau theory, and the resultant TM was dubbed the orthogonal magnetization (OM)~\cite{liang2017orthogonal}.
One common feature of the three systems in which TM was observed is that the antiferromagnetic ground state has
higher-rank CMM and the relevant spin Hamiltonian has spin-anisotropy arising from spin-orbit coupling.
Thus, to understand the fundamental origin of TM, the relation between the spin anisotropy and the complex magnetic structure
with higher-rank CMM should be clarified.

\begin{figure}[t]
	\centering
	\includegraphics[width=\columnwidth]{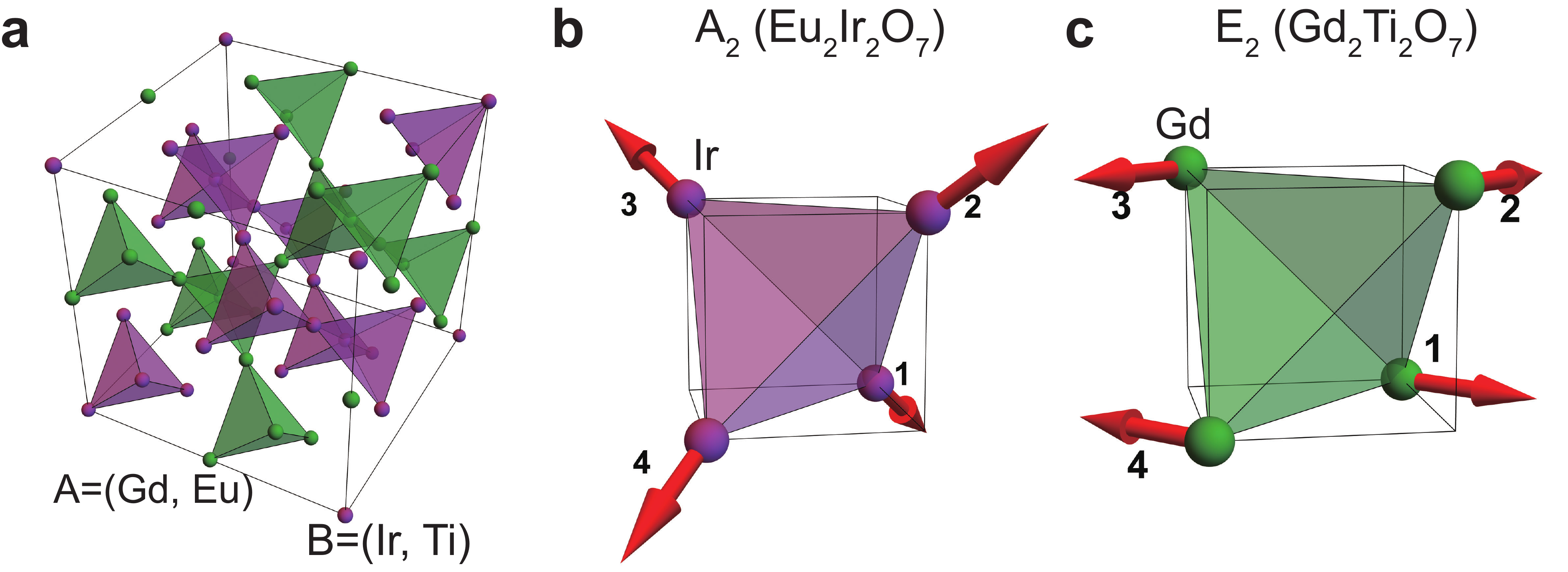}
	\caption{
(a) Structure of the pyrochlore lattice relevant to Gd$_2$Ti$_2$O$_7$ and Eu$_2$Ir$_2$O$_7$.
(b, c) Spin configurations of (b) A$_2$-octupole in Eu$_2$Ir$_2$O$_7$ and (c) E$_2$-dotriacontapole in Gd$_2$Ti$_2$O$_7$.
}
	\label{Fig1}
\end{figure}


In this Letter, we construct a general microscopic theory of TM.
First, through symmetry analysis, we derive the general symmetry condition to have TM.
Explicitly, we show that TM emerges only when every crystalline symmetry is broken, except for twofold antiunitary rotation $C_2T$, antiunitary mirror $\ma T$, and inversion $P$. 
Here, $C_2$, $\ma$, $T$ indicate two-fold rotation, mirror, and time-reversal symmetries, respectively.
Based on the symmetry, we further tabulate the information about whether TM is allowed or not under various field directions for all possible antiferromagnetic structures relevant to Mn$_3$Ir, CsMnBr$_3$, and pyrochlore systems including Gd$_2$Ti$_2$O$_7$ and Eu$_2$Ir$_2$O$_7$.

We also examine the microscopic origin of TM by studying the classical spin Hamiltonian on the pyrochlore lattice with spin anisotropy
represented by single-ion anisotropy (SIA), Dzyaloshinskii-Moriya interaction (DMI), and dipolar interaction (DI).
Depending on the nature of spin anisotropy, the antiferromagnetic ground state has distinct CMMs, and the degenerate ground state manifold (DGSM)
is either discrete or continuous under spin rotation. We find that when DGSM is discrete, TM always appears unless forbidden by symmetry.
On the other hand, when DGSM is continuous, TM is generally not allowed. 
However, when DGSM is constrained in easy planes by SIA, TM can appear when the magnetic field direction and spin configuration satisfy
certain geometric conditions. 
As a result of TM, we show that TM induces a unique transport phenomenon called anomalous planar Hall Effect (APHE)~\cite{battilomo2021anomalous}. 
Although we mainly focus on the pyrochlore lattice, our theory can be readily generalized to any antiferromagnets on any lattice system.

{\it Global symmetry constraints.|}
Let us first consider the symmetry constraint on the TM ($\Mb_\perp$) under $\Bb$.
First, we note that any $n$-fold rotation symmetry $C_n$ ($n=2,3,4,6$) along the direction of $\Bb$ prohibits nonzero TM because $\Mb_\perp$ is canceled by its rotated counterparts $\sum_{i=1}^{n-1} C_n^i \Mb_\perp$.
Similarly, a mirror symmetry $\sigma$ with the normal direction parallel to $\Bb$ also forbids the TM.
The only unitary symmetry compatible with nonzero TM is spatial inversion $P$.

In the case of antiunitary symmetries, there are two symmetries compatible with $\Mb_\perp\neq0$.
One is $C_2T$ symmetry whose rotation axis is perpendicular to $\Bb$.
In this case, $\Mb_\perp$ perpendicular to both $\Bb$ and the $C_2$ rotation axis can be nonzero.
The other is $\ma T$ symmetry whose mirror plane is parallel to $\Bb$.
Then, $\Mb_\perp$ can appear parallel to the mirror plane.
As the combination of $C_2T$ and $\ma T$ is just $P$, $\Mb_\perp$ can emerge even when both symmetries exist simultaneously. 
In summary, every symmetry except for $C_2T$, $\ma T$, and $P$ must be broken to have $\Mb_\perp\neq0$.

Using this symmetry condition,
one can judge whether $\Mb_\perp$ is forbidden or not in any antiferromagnetic (AFM) system under various field directions. 
In the case of AFM orders in the pyrochlore lattice with a tetrahedral magnetic unit cell shown in Fig.~\ref{Fig1},
the magnetic structures can be classified by using group theory, and the resulting irreducible representations (IRREPs) can be described in terms of CMMs~\cite{suzuki2017cluster,suzuki2019multipole} including $A_2$-octupole ($A_2$), 
$T_1$-octupoles ($T_{1x},T_{1y},T_{1z}$), $T_2$-octupoles ($T_{2x},T_{2y},T_{2z}$), and $E$-dotriacontapoles ($E_1,E_2$)~\cite{suzuki2017cluster,liang2017orthogonal,oh2018magnetic,suzuki2019multipole,kim2020strain,palmer2000order,elhajal2005ordering,gingras2014quantum}.
In the case of the $A_2$-octupole shown in Fig.~\ref{Fig1}, for example, its magnetic point group is $-4'3m'$ composed of an identity $I$, 3 two-fold rotations $C_2$, 8 threefold rotations $C_3'$, 6 antiunitary mirrors $\ma T$, and 6 four-fold antiunitary inversion $S_4T$. 
For $\Bb\parallel[001]$, every symmetry except $I$, $C_{2z}$, and two $\ma T$s is broken. 
Because there is $C_{2z}$, $\Mb_\perp=0$. 
On the other hand, when $\Bb\parallel[110]$, only $I$ and a $\ma T$ remain, thus $\Mb_\perp$ can be nonzero.
We extend this analysis to $D_{3h}$ point group relevant to CsMnBr$_3$~\cite{abarzhi1992spin} and to $O_h$ point group relevant to Mn$_3$Ir~\cite{zhang2017strong,suzuki2017cluster,tomeno1999magnetic,taylor2019epitaxial,chen2014anomalous,zhang2018spin}, as summarized in Appendix.

The analysis of the magnetic point group symmetry under $\Bb$ can also determine the direction of $\Mb_{\perp}$ and its general $\Bb$ dependence.
For instance, let us consider an AFM ordering with the magnetic point group $P$, which is described by the Hamiltonian $H(\{\Sb_a\})$
where $a$ is a sublattice index.
When $\Bb$ is applied, the symmetries in $P$ will be mostly broken but they still strongly constrain the spin canting directions.
More explicitly, for an element $O_{p} \in P$, we have 
\begin{align}
U(O_{p}) H(\{\Sb_a\},\Bb)U(O_{p})^{-1} = H(\{\Sb_a\},\Bb_p),
\label{eq:symmetryconstraint}
\end{align}
where $\Bb_p = O_p\Bb$ and $U(O_p)$ is the matrix representation of $O_p$. Namely, $O_p$ effectively changes the direction of $\Bb$ while keeping the spin structure.
For example, let us consider the $A_2$-octupole under $\Bb\parallel [110]$ again. 
Among the symmetries in $P$, $P_1 = \{ I, \sigma_{[1\bar10] }T \}$ indicates the symmetry that leaves $\Bb$ invariant.
Here $I$ denotes the identity and $\sigma_{[1\bar10]}$ is the mirror symmetry whose normal direction is along $[1\bar10]$. 
On the other hand, $P_2 = \{C_{2z}, \sigma_{[110]}T\}$ denotes the symmetries which invert the direction of $\Bb$. 
Here $\sigma_{[110]}$ is the mirror symmetry whose normal direction is along $[110]$. 
Applying $P_1$ and $P_2$ symmetries to the constraint equation in Eq.~(\ref{eq:symmetryconstraint}), we obtain
$\Mb_\perp \propto \left[b B^2 + O(B^4) \right]  \hat z $ with a constant $b$.
A similar analysis can also be applied to other CMMs.
In the case of $E_2$-dotriacontapole under $\Bb \parallel [111]$, we find that $P_1 = \{I\}$ leaves $\Bb$ invariant 
while $P_2 = \{\ma_{1\bar10}\}$, inverts the $\Bb$ direction,
which gives $\Mb_\perp= (bB^2+...) \hat e_{1\bar10} + (dB+fB^3+...) \hat e_{11\bar2}$
where $b,~d,~f$ are constants.
Detailed $\Bb$ dependence of TM is determined by microscopic spin interactions as discussed below.
The cases of $E_1$ and $T_{2y}$ CMMs under $\Bb \parallel [111]$ are further analyzed in Appendix.

{\it Microscopic Hamiltonian.|}
The classical Heisenberg antiferromagnet on the pyrochlore lattice has macroscopically degenerate ground states \cite{gardner2010magnetic,moessner1998properties}. 
Under a magnetic field $\Bb$, the Hamiltonian can be written as
\begin{align}
	H_{0} =& H_J + H_B =J\sum_{\langle ab \rangle } \Sb_a \cdot \Sb_b - \sum_a \Bb \cdot \Sb_a,\n
	=& 8 J N_c \Mb^2 - 4 N_c \Bb \cdot \Mb - \f{J N_c}{2}\sum_{a=1}^4 \Sb_a^2,
\end{align}
where $H_J$ with $J>0$ indicates the isotropic antiferromagnetic exchange interaction
between nearest-neighboring spins, and $H_B$ is the Zeeman coupling.
$N_c$ is the number of tetrahedral unit cells, $\Mb = \f{1}{4}\sum_{a=1}^4 \Sb_a$ is the average magnetization
of the four spins in a tetrahedron. 
From $\Sb_a^2=1$, we obtain $H_0 = 8J N_c (\Mb-\f{\Bb}{4J})^2$. 
Then, the minimum energy condition gives $\Mb = \f{\Bb}{4J}$. 
Namely, TM does not appear when spin anisotropy is absent.

\begin{figure}
	\centering
	\includegraphics[width=\columnwidth]{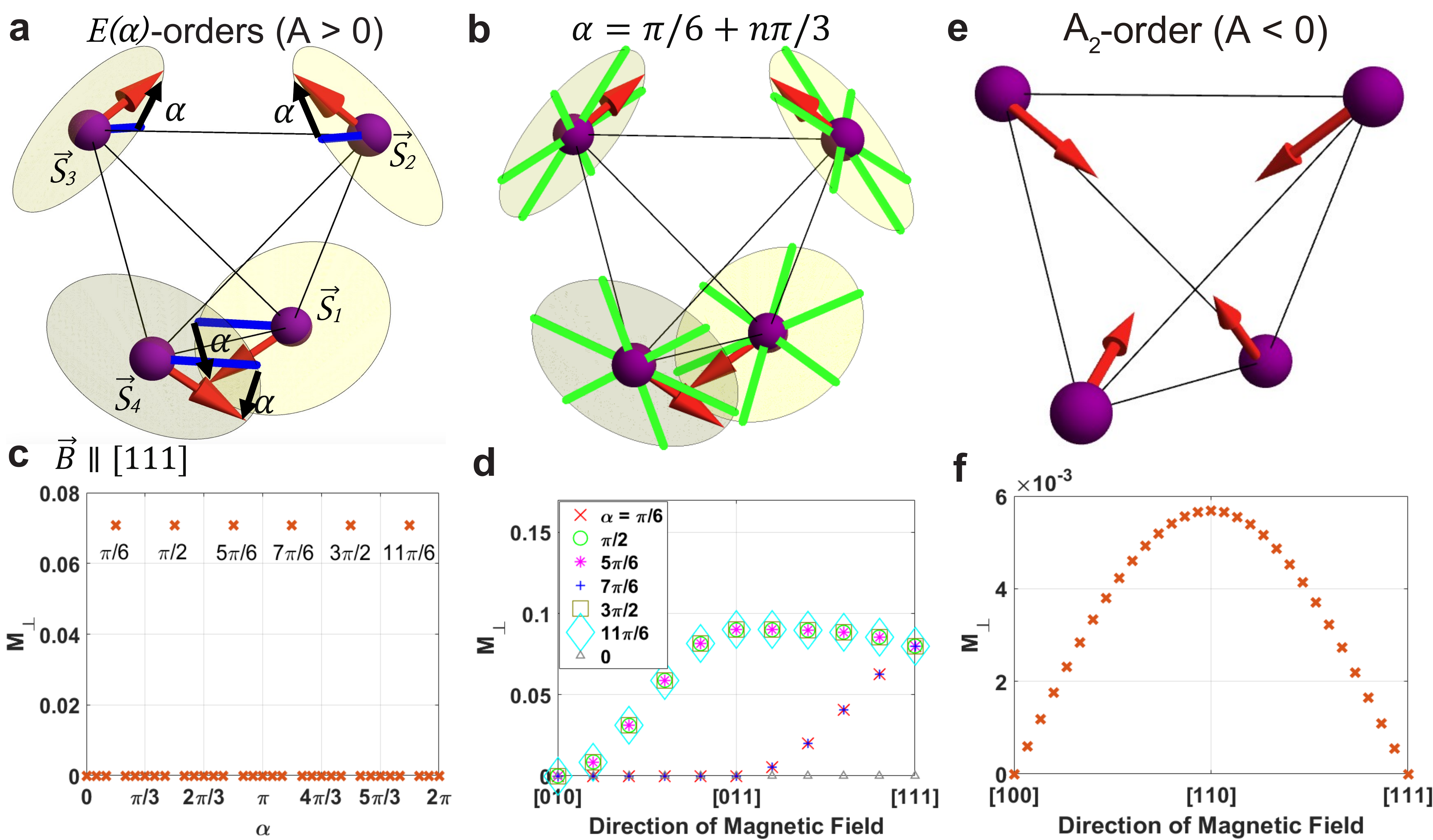}
	\caption{
		(a) $\hat{E}(\alpha)$-order when $A>0$. The spins (red arrows) are lying on their easy planes (yellow planes).
		(b) Green lines denote the spin directions of $E(\pi/6+n\pi/3)$-orders ($n=0,\cdots,5$). 
		(c) $\Mb_{\perp}$ for $\hat{E}(\alpha)$-order as a function of $\alpha$ when $\Bb\parallel [111]$. 
		$\Mb_{\perp}\neq0$ only at $\alpha = \pi/6+n\pi/3$. 
		(d) $\Mb_{\perp}$ for $\hat{E}(\alpha)$-order with various $\alpha$ computed by changing $\Bb$ from $[010]$ to $[011]$, and to $[111]$. 
		(e) $A_2$-order when $A<0$.
		(f) $\Mb_{\perp}$ for $A_2$-order computed by changing $\Bb$ from $[100]$ to $[110]$, and to $[111]$.  
		In (c,d,f), we choose $|A|/J=B/J=1$.
	}
	\label{Fig2}
\end{figure}

{\it Single-ion anisotropy (SIA).|}
Let us consider $H_1\equiv H_0+H_{SIA}$ that includes the SIA,
$H_{SIA}=A\sum_a (\Sb_a\cdot\mathbf{n}_a)^2$. 
When $A>0$, $H_{SIA}$ forces $\Sb_a$ to lie on its easy plane
on which $\Sb_a \cdot \mathbf{n}_a = 0$ is satisfied [see Fig.~\ref{Fig2}a].
The energy minimum condition of $H_1$ is
\begin{align}
	\Mb = \f{\Bb}{4J},\quad \Sb_a\cdot \mathbf{n}_{a} = 0~(a=1,2,3,4). \label{Emin_SIA}
\end{align}
When $\Bb=0$, the ground state is antiferromagnetic with either $E$-dotriacontapole or $T_2$-octupole, 
in which all spins are lying on their easy-planes.
As the $E$-dotriacontapole belongs to a two-dimensional (2D) IRREP,
it is composed of two basis states called the $\hat{E}_1$ and $\hat{E}_2$-orders.
Similarly, the $T_2$-octupole belonging to a three-dimensional (3D) IRREP 
is composed of three basis states, called the $\hat{T}_{2x}$, $\hat{T}_{2y}$, and $\hat{T}_{2z}$-orders [see Appendix].

More specifically, in the $\hat{E}_1$-order,
the four spins $\Sb_{a=1,2,3,4}$ in a unit cell are aligned along the directions $\hat x_1 = [1\bar10]$, $\hat x_2 = [110]$, $\hat x_3 = [\bar1\bar10]$, $\hat x_4 = [\bar110]$, respectively, while for the $\hat{E}_2$-order, the spins are along $\hat y_1 = [11\bar2]$, $\hat y_2 = [1\bar12]$, $\hat y_3 = [\bar112]$, $\hat y_4 = [\bar1\bar1\bar2]$, respectively.
Then a general $E$-dotriacontapole order can be represented by $\hat{E}(\alpha) = \hat{E}_1\sin\alpha+\hat{E}_2\cos\alpha$,
which spans a continuous DGSM parametrized by $0\leq\alpha\leq2\pi$  [see Fig.~\ref{Fig2}a]. 
As $\alpha$ varies, the spins continuously rotate on their easy planes.
Similar to $(\hat{E}_1, \hat{E}_2)$-orders, $(\hat{E}(\alpha=\pi/6), \hat{T}_{2x})$-orders,
$(\hat{E}(\alpha=5\pi/6), \hat{T}_{2y})$-orders, and $(\hat{E}(\alpha=\pi/2), \hat{T}_{2z})$-orders
form pairs of basis states which span continuous DGSM where spins are lying on their easy planes.

When $\Bb\neq0$, the energy minimum condition in Eq.~(\ref{Emin_SIA}) 
is satisfied in most cases, thus TM vanishes. But there are a few exceptional cases with nonzero TM. 
For example, for a given $\hat{E}(\alpha)$ order at $\Bb=0$, the spin configuration at small $\Bb$ can be parametrized as 
\begin{align}
	\Sb_a =& \cos\ta_a\left[\cos(\alpha+\phi_a)\hat x_a 
	+ \sin(\alpha+\phi_a)\hat y_a\right] - \sin\ta_a\hat z_a, \label{angles}
\end{align}
where $\phi_a$ ($\ta_a$) indicates the rotation within (away from) the easy-plane of $\Sb_a$ due to $\Bb\neq0$. 
At small $\Bb$, we expand $\Sb_a$ up to the first order of  ($\ta_a$,$\phi_a$) and put it in Eq.~(\ref{Emin_SIA}), 
which gives $\Sb_a\cdot \mathbf{n}_a = - \ta_a = 0$, $M_x=\frac{1}{4\sqrt{6}}\left[(\cos\A-\sqrt{3} \sin\A)(\phi_1 +\phi_2 -\phi_3 - \phi_4)\right]$,
$M_y=\frac{1}{4\sqrt{6}}\left[(\cos\A+\sqrt{3} \sin\A)(\phi_1 -\phi_2 +\phi_3 - \phi_4)\right]$,
$M_z=\frac{-1}{2\sqrt{6}}\left[\cos\A(\phi_1 -\phi_2 -\phi_3 + \phi_4)\right]$.
Note that when $\tan\alpha=1/\sqrt{3}$, $M_x=0$. Then, $M_x=B_x/(4J)$ in Eq.~(\ref{Emin_SIA}) cannot be satisfied
if $B_x\neq0$. 
Similar situations occur when $\tan\alpha=-1/\sqrt{3}$ and $B_y\neq 0$, or $\cos\alpha=0$ and $B_z\neq0$.	

Interestingly, these are exactly the conditions to have nonzero TM [see Fig.~\ref{Fig2}b].
For instance, for the $\hat{E}(\alpha=\pi/6, 7\pi/6)$ order with $\tan\alpha=1/\sqrt{3}$,
when $\Bb\parallel[100]$, the projection of $\Bb$ onto the easy plane of each spin is parallel to the corresponding spin direction, thus $\Bb$ cannot rotate each spin within its easy plane.
Instead, $\Bb$ forces the spins to move away from their easy planes, which makes the energy minimum condition in Eq.~(\ref{Emin_SIA}) to be violated and induces nonzero TM.
Similar situations happen for $\hat{E}(5\pi/6, 11\pi/6)$ order with $B_y\neq0$,
and $\hat{E}(\pi/2, 3\pi/2)$ order with $B_z\neq0$.

The spin configuration with nonzero TM can be obtained by the stationary condition
$\pa H_1 /\pa \ta_a = \pa H_1/\pa \phi_a = 0$. 
For instance, for $\hat{E}(\pi/2)$ order under $\Bb\parallel[111]$ described in Fig.~\ref{Fig2}b, the stationary condition gives
$\ta_1 = \ta_4 = -\ta_2 = -\ta_3  = \f{B}{6A+4J}$, $\phi_1 = \phi_4 = 0$, $\phi_2 = -\phi_3 = -\f{B}{\sqrt6 J}$,
from which we obtain $\Mb_{\perp} = \f{\sqrt2 }{4(2+3A/J)}(\f{A}{J})(\f{B}{J})\hat e_{11\bar2}$. 
We note that as $\theta_{1,2,3,4}$ are nonzero, all spins move away from their easy planes.
In Fig.~\ref{Fig2}c, we compute $\Mb_{\perp}$ for $\hat{E}(\alpha)$ order under $\Bb \parallel [111]$ varying $\alpha$. 
In Fig.~\ref{Fig2}d, we plot $\Mb_{\perp}$ for various $\hat{E}(\alpha)$-orders by continuously rotating $\Bb$ from $[010]$ to $[011]$, and then to $[111]$ in sequence. 
$\Mb_{\perp}$ becomes nonzero only when the special conditions between $\Bb$ and $\alpha$ described above are satisfied. [See Appendix for further discussions.]

When $A<0$, on the other hand, each spin $\Sb_a$ aligns along its easy axis direction $\mathbf{n}_a$,
leading to the all-in all-out ground state with an $A_2$-octupolar moment shown in Fig.~\ref{Fig2}e.
Two degenerate ground states, all-in or all-out state, related by time-reversal symmetry form a discrete manifold
in which the states are separated by an energy barrier, contrary to the $A>0$ case.
In this situation, we find that TM can generally appear unless it is forbidden by symmetry.
We compute the TM by changing $\Bb$ from $[010]$ to $[101]$, and then to $[111]$ continuously, and represent the result in Fig.~\ref{Fig2}f. Note that considering symmetry, TM vanishes for $\Bb \parallel [001]$ and $[111]$. For other directions, TM is nonzero and exhibits $|\Mb_\perp| \propto (A/J)(B/J)^2$ consistent with magnetic space group analysis. 
All these results are further confirmed by numerically solving $H_1$ using mean-field theory [see Appendix].

\begin{figure}
	\centering
	\includegraphics[width=\columnwidth]{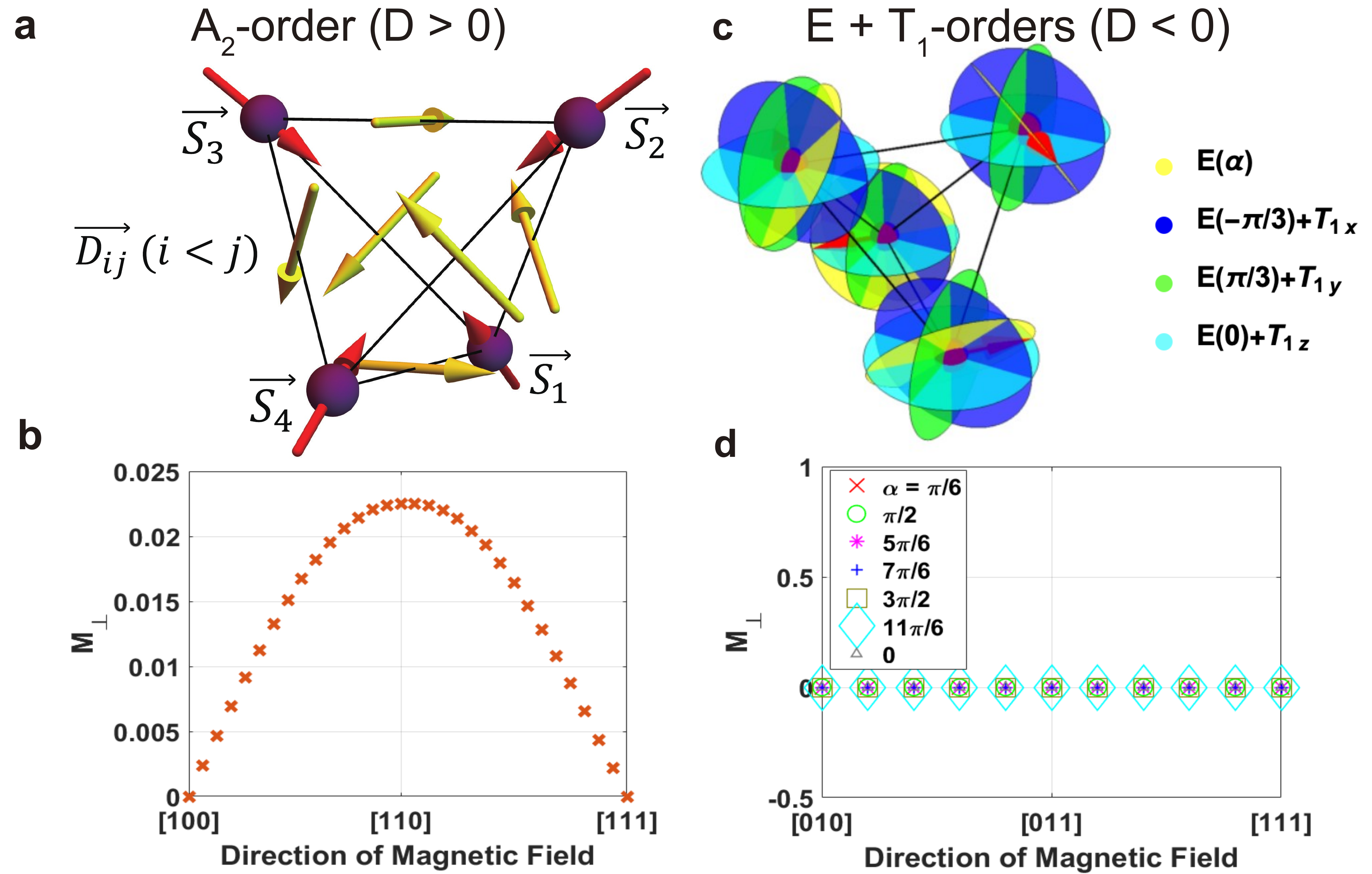}
	\caption{
		(a) $A_2$-order when $D>0$, where all spins (red arrows) are perpendicular to the surrounding DM vectors (yellow arrows).
		(b) $\Mb_{\perp}$ for $A_2$-order computed by changing $\Bb$ from $[100]$ to $[110]$ and to $[111]$. 
		(c) Schematic description of continuous DGSM when $D<0$. There are four distinct planes; yellow planes are for $E(\A)$-order, blue planes are for $E(-\pi/3)+T_{1x}$-order, green planes are for $E(\pi/3)+T_{1y}$-order, and cyan planes are for $E(0)+T_{1z}$-order. 
		(d) $\Mb_{\perp}$ for $\hat{E}(\alpha)$-order with various $\alpha$ computed by changing $\Bb$ from $[010]$ to $[011]$ and to $[111]$.  In (b,d), we assume $|D|/J = B/J = 1$. 
	}	\label{Fig3}
\end{figure}

{\it Dzyaloshinskii-Moriya interaction (DMI).|}
Next, we consider $H_2\equiv H_{0}+H_{DMI}$ that includes the DMI,
$H_{DMI}=D \sum_{\langle ab \rangle} \hat \Db_{ab}\cdot (\Sb_a\times \Sb_b)$.
Generally, DMI forces two spins $\Sb_a$ and $\Sb_b$ to lie in their planes perpendicular to the DM vector $\hat \Db_{ab}$ so that $\Sb_a \times \Sb_b$ is anti-parallel (parallel) to $\hat\Db_{ab}$ when $D>0$ $(D<0)$.

Let us first consider $D>0$ case~\cite{elhajal2005ordering}. In the pyrochlore lattice,
DMI forces each spin to be perpendicular to its six neighboring DM vectors,
and the intersection between the planes normal to those DM vectors is uniquely determined, which leads to the $A_2$-order as shown in Fig.~\ref{Fig3}a.
As in the case of SIA with $A<0$, since DGSM is discrete, TM can generally arise unless prohibited by symmetry. 
For example, in Fig.~\ref{Fig3}b, we compute the TM by changing $\Bb$ from $[100]$ to $[110]$, and then to $[111]$. When $\Bb \parallel [100]$ and $[111]$, TM vanishes because of rotation symmetries. Otherwise, TM is nonzero. From the stationary condition of $H_2$, we obtain $|\Mb_{\perp}|\propto (\f{D}{J})(\f{B^2}{J^2})$.
[See Appendix for more details.]

On the other hand, when $D<0$, the relative angle between neighboring spins should be inverted compared to $D>0$ case to minimize the energy. To find the ground state for $D<0$, we rewrite $H_2$ by adding some constants as
\begin{align}
	H_2 =& -12D (\sum_a \Sb_a \cdot \hat v_a/4)^2 - 8D \sum_{r=1}^3[(\sum_{a} \Sb_a \cdot \mathbf{T}_{a}^{r}/4)^2] \n&+8(J-D/2)(\Mb-\f{\Bb}{4(J-D/2)})^2,
\end{align}
where $\hat v_a$ is the unit vector along the local $z$-axis of $\Sb_a$, and $\mathbf{T}_{a}^{r}$ ($r=1,2,3$) indicates the spin direction relevant to $T_2$ octupolar ordering. The explicit forms of $\hat v_a$ and $\mathbf{T}_{a}^{r}$ are given in Appendix.
Since $D<0$, all the coefficients of squared terms in $H_2$ are positive, thus $H_2$ can be minimized when the following seven equations are satisfied, 
\begin{align}
	\Mb = \f{\Bb}{4J-2D},~\sum_{a=1}^{4} \Sb_a \cdot \hat v_a = 0,~\sum_{a=1}^{4} \Sb_a \cdot \mathbf{T}_a^r = 0. \label{Emin_DM}
\end{align}

When $\Bb = 0$, one can show that $(\hat{E}_1, \hat{E}_2)$-orders span the continuous DGSM as in the case of SIA with $A>0$. 
Similarly, $(\hat{E}(\alpha=0), \hat{T}_{1z})$-orders,
$(\hat{E}(\alpha=\pi/3), \hat{T}_{1y})$-orders, and $(\hat{E}(\alpha=-\pi/3), \hat{T}_{1x})$-orders
form pairs of basis states which span continuous DGSM where spins are lying on the $xy$, $zx$, and $yz$ planes, respectively. [See Fig.~\ref{Fig3}c.]

Since DGSM is continuous, one can generally expect TM to be vanishing. To check the possible exceptional situations as in the SIA case with $A>0$, let us consider $\hat{E}(\alpha)$ order at $\Bb=0$ and examine the spin configuration at small $\Bb$ by introducing angular variation $(\ta_a,\phi_a)$ as in Eq.~(\ref{angles}).
Plugging the parametrized form of spins in Eq.~(\ref{angles}) into Eq.~(\ref{Emin_DM}), we obtain, up to the linear order in $B=|\Bb|$,
$\ta_a = -3 \hat z_a \cdot \f{\Bb}{4(J-D/2)}$, $\phi_a = q \f{B}{4(J-D/2)}$
where $q$ is an arbitrary constant.
Contrary to the case of SIA with $A>0$ in which $\ta_a=0$ is always required to minimize the SIA term irrespective of $\Bb$, in the DMI case with $D<0$, both $\ta_a$ and $\phi_a$ can continuously vary under $\Bb$ while the energy minimum condition is satisfied. As spins can rotate continuously in three-dimensional space under $\Bb$ while satisfying Eq.~(\ref{Emin_DM}), TM does not appear. This is generally true for arbitrary $E(\A)$ under arbitrary $\Bb$, as shown in Fig.~\ref{Fig3}d.
The same results can be obtained from the stationary conditions
$\pa H_2 /\pa \ta_a = \pa H_2/\pa \phi_a = 0$. 
All these results can be further confirmed by numerical mean-field calculation of $H_2$ [see Appendix].
Also, other $\hat{E}+\hat{T}_1$-type ground states with continuous DGSM exhibit similar behaviors as discussed in Appendix.

\begin{figure}[b]
	\centering
	\includegraphics[width=\columnwidth]{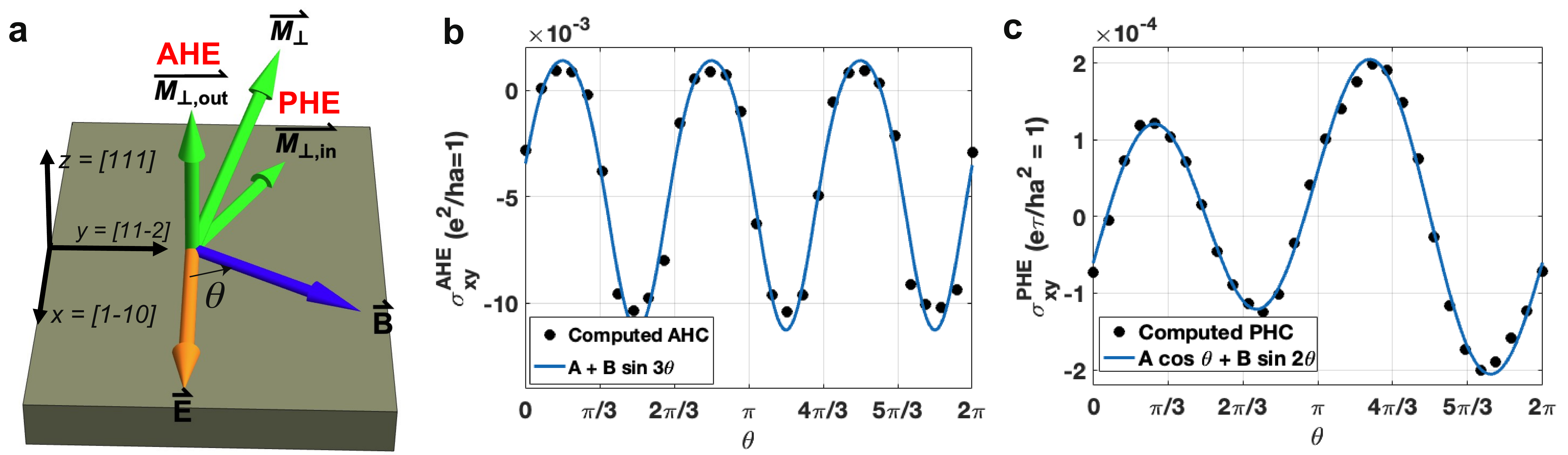}
	\caption{
		(a) Schematic description of TM and APHE. 
		(b) The computed AHC (black dots) and its fitting (blue line) by $ \beta_0 + \beta_2\sin 3\ta$. 
		(c) The computed PHC (black dots) and its fitting (blue line) by $\gamma_1\cos\ta + \delta_1\sin2\ta$.
	}
	\label{Fig4}
\end{figure}

{\it Anomalous Planar Hall Effect (APHE).|}
In metallic antiferromagnets with CMMs, TM can induce
APHE~\cite{battilomo2021anomalous}, i.e. simultaneous appearance of anomalous Hall effect (AHE)
and planar Hall effect (PHE)~\cite{nandy2017chiral,zheng2020origin, ky1966plane}.
Such a thing is possible because an applied in-plane $\Bb$ can generate 
both in-plane $\Mb_{\perp,in}$ and out-of-plane $\Mb_{\perp,out}$ TM,
which give PHE and AHE, respectively. [See Fig.~\ref{Fig4}a.] 

Motivated by the recent experimental observation of AHE and PHE in pyrochlore iridates~\cite{kim2020strain,li2021correlated},
we examine APHE in this system with $A_2$-order. 
Assuming $\hat x \parallel[1\bar10]$, $\hat y \parallel[11\bar2]$, and $\hat z \parallel [111]$,
we apply an electric field $\Eb\parallel\hat x$, and rotate $\Bb$ within $\hat x\hat y$-plane.
Considering the symmetry of pyrochlore lattice, we find
$M_{\perp,out} = (a_0 + a_1 \cos 3\ta + b_1 \sin 3\ta)\hat z$, $M_{\perp,in} = (c_1 \cos 3\ta)\hat p$
where $a_0, a_1, b_1,$ and $c_1$ are constants, and $\hat p = (-\sin\ta,\cos\ta,0)$ \red{} is the in-plane unit vector perpendicular to $\Bb$. [See Fig.~\ref{Fig4}a.]
Using the phenomenological model for anomalous Hall conductivity (AHC) and planar Hall conductivity (PHC)~\cite{wang2020antisymmetric,nandy2018berry} given by
$\ma_{xy}^{AHE} = \ma_0 M_{\perp,out}$, $\ma_{xy}^{PHE} = \ma_1 B_xB_y + \ma_2 (B_xM_{\perp,y}+B_yM_{\perp,x}) +  \ma_3 M_{\perp,x}M_{\perp,y}$, we obtain
\begin{align}
	\ma_{xy}^{AHE} \propto& \beta_0 + \beta_1\cos3\ta + \beta_2 \sin3\ta,\n
	\ma_{xy}^{PHE} \propto& \gamma_1\cos\ta + \gamma_2 \cos5\ta \n&+ \delta_1\sin2\ta +\delta_2\sin4\ta+\delta_3\sin8\ta, \label{eqPM3}
\end{align}
where $\beta_{0,1,2}$, $\gamma_{1,2}$, and $\delta_{1,2,3}$ are constants.
We note that $\gamma_1, \gamma_2$ terms come from $\ma_2$ term in $\ma_{xy}^{PHE}$  
while $\delta_1,\delta_2,\delta_3$ come from $\ma_3$ term in $\ma_{xy}^{PHE}$.

To confirm the prediction of the above phenomenological theory,
we perform self-consistent mean-field calculations of the Hubbard model describing pyrochlore iridates with $A_2$-order~\cite{witczak2013pyrochlore,oh2018magnetic},
and numerically compute the AHC~\cite{nagaosa2010anomalous} and PHC~\cite{nandy2017chiral}. [See Appendix for details.]
For AHC, we consider only the intrinsic Berry curvature contribution while for PHC, we assume constant relaxation time.
The resulting AHC (PHC) is plotted using black dots in Fig.~\ref{Fig4}b (Fig.~\ref{Fig4}c) which
can be fitted by $\ma_{xy}^{AHE} \propto \beta_0 + \beta_2\sin 3\ta$ and $\ma_{xy}^{PHE} \propto \gamma_1\cos\ta + \delta_1 \sin 2\ta$, respectively, consistent with Eq.~(\ref{eqPM3}). 
We note that experimental data can contain additional terms due to the presence of rare-earth ions and strain, etc.~\cite{li2021correlated}. 
As APHE can probe multipolar AFM structures through its relation with TM, 
it can be further applied to the systems where conventional methods like neutron scattering do not work~\cite{glazkov2005single,glazkov2006observation,glazkov2007single}.

{\it Discussion.|}
To conclude, we construct a general microscopic theory of TM and identify the symmetry condition to have nonzero TM.
When DGSM is discrete, TM is generally allowed unless forbidden by symmetry.
On the other hand, when DGSM is continuous, TM generally vanishes except the SIA case with $A>0$.
We have also analyzed the cases with dipolar interactions and obtained nonzero TM due to the discrete DGSM [see Appendix].

Our theory can successfully explain the experimental data for CsMnBr$_{3}$, Gd$_2$Ti$_2$O$_7$, Eu$_2$Ir$_2$O$_7$ as shown in Appendix.
Especially, in the case of 
Eu$_2$Ir$_2$O$_7$, the TM induced by spin canting shows the same behaviors as the OM from phenomenological Landau theory. Thus we think that the spin canting contribution to OM cannot be ruled out to explain the measured data.

\section*{Acknowledgment}
	T.O., S.P., and B.J.Y. were supported by the Institute for Basic Science in Korea (Grant No. IBS-R009-D1), 
	Samsung  Science and Technology Foundation under Project Number SSTF-BA2002-06,
	and the National Research Foundation of Korea (NRF) grant funded by the Korea government (MSIT) (No. 2021R1A2C4002773, and No. NRF-2021R1A5A1032996).

\clearpage

\section*{Appendices to "Theory of transverse magnetization in spin-orbit coupled antiferromagnets"}

\appendix
\tableofcontents
\renewcommand{\thefigure}{S\arabic{figure}}

\renewcommand{\thetable}{S\arabic{table}}
\renewcommand{\arraystretch}{1.5}
\setcounter{figure}{0}
\setcounter{equation}{0}
\setlength{\tabcolsep}{3pt}

	\section{The symmetry condition for TM \label{Ssec1}}
	
	Let us consider the $A_2$-order on the pyrochlore lattice under $\Bb\parallel [110]$. 
	Among the magnetic point group symmetries, $P_1 = \{ I, \sigma_{[1\bar10] }T \}$ indicates the symmetry that leaves $\Bb$ invariant.
	Here $I$ denotes the identity and $\sigma_{[1\bar10]}$ is the mirror symmetry whose normal direction is along $[1\bar10]$. 
	On the other hand, $P_2 = \{C_{2z}, \sigma_{[110]}T\}$ denotes the symmetries which invert the direction of $\Bb$. 
	Here $\sigma_{[110]}$ is the mirror symmetry whose normal direction is along $[110]$. 
	$P_1$ and $P_2$ symmetries give the following relations between $\Dt \Sb_{i}(\Bb)$:
	\begin{align}
		&\Dt \Sb_{1,1\bar1 0} (\Bb) = \Dt \Sb_{4,1\bar10}(\Bb) = 0, \n
		&\Dt \Sb_{1,z} (\Bb) = \Dt \Sb_{4,z}(-\Bb), \n
		&\Dt \Sb_{2,1\bar10}(\Bb) =-\Dt \Sb_{3,1\bar10}(\Bb) \n&= \Dt\Sb_{2,1\bar10}(-\Bb) = -\Dt\Sb_{3,1\bar10}(-\Bb), \n
		&\Dt \Sb_{2,z}(\Bb) =\Dt \Sb_{3,z}(\Bb) \n&= \Dt\Sb_{2,z}(-\Bb) = \Dt\Sb_{3,z}(-\Bb).
	\end{align}
	Accordingly, the transverse spin change takes the following form
	\begin{align}
		\Dt \Sb_{1\perp} =& (a_1B+b_1B^2+c_1B^3+...)\hat z, \n
		\Dt \Sb_{2\perp} =& (b_2B^2+...)\hat z + (b_3 B^2 + ...) \hat e_{1\bar10}, \n
		\Dt \Sb_{3\perp} =& (b_2B^2+...)\hat z - (b_3 B^2 + ...) \hat e_{1\bar10},\n
		\Dt \Sb_{4\perp} =& (-a_1B+b_1B^2-c_1B^3+...)\hat z,
	\end{align}
	where $a_1, b_1, b_2, b_3$, and $c_1$ are constants, and $\hat e_{1\bar10} = (\hat x - \hat y)/\sqrt2$. 
	Note that as the initial directions of $\Sb_2$ and $\Sb_3$ are perpendicular to $\Bb$, 
	their transverse components only have even powers of $\Bb$. 
	Hence, we finally obtain the spin TM as
	\begin{align}
		\Mb_{\perp} \propto \left[b B^2 + O(B^4) \right]  \hat z \label{eqTM1}.
	\end{align}
	
	A similar analysis can be applied to any CMM.
	In the case of $E_2$-dotriacontapole under $\Bb \parallel [111]$, we find that $P_1 = \{I\}$ leaves $\Bb$ invariant 
	while $P_2 = \{\ma_{1\bar10}\}$, inverts the $\Bb$ direction. 
	The constraints by $P_1$ and $P_2$ are
	\begin{align}
		&	\Dt \Sb_{1/4,1\bar10} (\Bb) = \Dt \Sb_{1/4,1\bar10} (-\Bb), \n
		&	\Dt \Sb_{1/4,11\bar2} (\Bb) = -\Dt \Sb_{1/4,11\bar2} (-\Bb), \n
		&	\Dt \Sb_{2,1\bar10} (\Bb) = \Dt \Sb_{3,1\bar10} (-\Bb), \n
		&   \Dt \Sb_{2,11\bar2} (\Bb) = -\Dt \Sb_{3,11\bar2} (-\Bb),
	\end{align}
	which give
	\begin{align}
		\Dt\Sb_{1,\perp} =& (b_1B^2 + ...)\hat e_{1\bar10} + (d_1B+f_1B^3 + ...)\hat e_{11\bar2}, \n
		\Dt\Sb_{2,\perp} =& (a_2B+b_2B^2+c_2B^3)\hat e_{1\bar10} \n&+ (d_2B+e_2B^2+f_2B^3)\hat e_{11\bar2}, \n
		\Dt\Sb_{3,\perp} =& (-a_2B+b_2B^2-c_2B^3)\hat e_{1\bar10} \n&+ (d_2B-e_2B^2+f_2B^3)\hat e_{11\bar2}, \n
		\Dt\Sb_{4,\perp} =& (b_4B^2 + ...)\hat e_{1\bar10} + (d_4B+f_4B^3 + ...)\hat e_{11\bar2} .
	\end{align}
	Hence, the spin TM is
	\begin{align}
		\Mb_{\perp} = (bB^2+...) \hat e_{1\bar10} + (dB+fB^3+...) \hat e_{11\bar2}. \label{eqTM2}
	\end{align}
	This shows that $\Mb_{\perp} \parallel [1\bar10]$ is even in $\Bb$ while $\Mb_{\perp} \parallel [11\bar2]$ is odd in $\Bb$.
	In SI, we have also considered $E_1$ and $T_{2y}$ CMMs under $\Bb\parallel[111]$.
	Here, we explain the other cases, $E_1$ under $\Bb \parallel [111]$ and $T_{2y}$ under $\Bb \parallel [111]$. Let $G_M$ is the magnetic point group under the magnetic order. Then, please note that $P_1$ is the set of $G_M$ elements which keeps $\Bb$ direction, while $P_2$ is the set of $G_M$ elements which reverses $\Bb$ direction.
	
	Next, let us think of $E_1$ under $\Bb \parallel [111]$. Then, $P_1 = \{I,\ma_{[1\bar10]}T\}$, $P_2 = \phi$. The condition is given by
	\begin{align}
		&	\Dt\Sb_{1/4,1\bar10}(\Bb) = 0, \n
		&	\Dt\Sb_{2,1\bar10}(\Bb) =  - \Dt\Sb_{3,1\bar10}(\Bb), \n
		&	\Dt\Sb_{2,11\bar2}(\Bb) =  \Dt\Sb_{3,11\bar2}(\Bb). 
	\end{align}
	Hence, each spin change is
	\begin{align}
		\Dt\Sb_{1\perp} =& s_1(B)\hat e_{11\bar2},\n
		\Dt\Sb_{2\perp} =& t_1(B)\hat e_{1\bar10} + s_2(B)\hat e_{11\bar2},\n
		\Dt\Sb_{3\perp} =& -t_1(B)\hat e_{1\bar10} + s_2(B)\hat e_{11\bar2},\n
		\Dt\Sb_{4\perp} =& s_4(B)\hat e_{11\bar2},
	\end{align}
	where $s_i,t_i$ are the polynomials of $B$. The TM is
	\begin{align}
		\Mb_{\perp} = s(B)\hat e_{11\bar2},
	\end{align}
	where $s(B) = aB+bB^2+cB^3+...$ is polynomial of $B$.
	
	Lastly, let us think of $T_{2y}$ under $\Bb \parallel [111]$, Then, $P_1 = \{I\}$, $P_2 = \{\ma_{[10\bar1 }\}$. The condition is given by,
	\begin{align}
		&	\Dt \Sb_{1/3,10\bar1}(\Bb) = \Dt\Sb_{1/3,10\bar1}(-\Bb), \n
		&	\Dt \Sb_{1/3,1\bar21}(\Bb) = -\Dt\Sb_{1/3,1\bar21}(-\Bb), \n
		&	\Dt \Sb_{2,10\bar1}(\Bb) = \Dt\Sb_{4,10\bar1}(-\Bb), \n
		&	\Dt \Sb_{2,1\bar21}(\Bb) = -\Dt\Sb_{4,1\bar21}(-\Bb).
	\end{align}
	Each spin change becomes
	\begin{align}
		\Dt\Sb_{1\perp} =& (b_1B^2+...)\hat e_{10\bar1} + (d_1B+f_1B^3+...)\hat e_{1\bar21},\n
		\Dt\Sb_{2\perp} =& (a_2B+b_2B^2+c_2B^3+...)\hat e_{10\bar1} \n&+ (d_2B+e_2B^2+f_2B^3+...)\hat e_{1\bar21},\n
		\Dt\Sb_{3\perp} =& (b_3B^2+...)\hat e_{10\bar1} + (d_3B+f_3B^3+...)\hat e_{1\bar21},\n
		\Dt\Sb_{4\perp} =& (-a_2B+b_2B^2-c_2B^3+...)\hat e_{10\bar1} \n&+ (d_2B-e_2B^2+f_2B^3+...)\hat e_{1\bar21},
	\end{align}
	Thus, the TM is
	\begin{align}
		\Mb_{\perp} = (bB^2+...) \hat e_{10\bar1} + (dB+fB^3+...)\hat e_{1\bar21}.
	\end{align}
	
	In fact, we can do the same procedure for all symmetries in the MPG. However, because any symmetries other than $P_1$ and $P_2$ change the field direction, they does not give further physical meanings to the field dependence of TM. For example, under AIAO and $\Bb \parallel [110]$, recall that from $P_1 = \{I, \ma_{1\bar10}T\}$ and $P_2 = \{ C_{2z},\ma_{110}T \}$, we have the following form of transverse spin change.
	\begin{align}
		\Dt S_{1\perp}(\Bb) =& g_1(\Bb)\hat z,\n
		\Dt S_{2\perp}(\Bb) =& f_2^{even}(\Bb)\hat e_{1\bar10} + g_2^{even}(\Bb) \hat z,\n
		\Dt S_{3\perp}(\Bb) =& -f_2^{even}(\Bb)\hat e_{1\bar10} + g_2^{even}(\Bb) \hat z,\n
		\Dt S_{4\perp}(\Bb) =& g_1(-\Bb)\hat z,
	\end{align} 
	where $f^{even}(\Bb) = f^{even}(-\Bb)$. So the form of TM is $\Dt S_{tot,\perp}(\Bb) = g^{even}(\Bb)\hat z$.
	
	We can apply $C_3\notin P_1,P_2$ which is in the MPG of AIAO. Then, the condition is
	\begin{align}
		&\Dt S_{1,1\bar10}(\Bb) = \Dt S_{1,01\bar1}(C_3\Bb),\n
		&\Dt S_{1,z}(\Bb) = \Dt S_{1,x}(C_3\Bb),\n
		&\Dt S_{2,1\bar10}(\Bb) = \Dt S_{3,01\bar1}(C_3\Bb),\n
		&\Dt S_{2,z}(\Bb) = \Dt S_{3,x}(C_3\Bb),\n
		&\Dt S_{3,1\bar10}(\Bb) = \Dt S_{4,01\bar1}(C_3\Bb),\n
		&\Dt S_{3,z}(\Bb) = \Dt S_{4,x}(C_3\Bb),\n
		&\Dt S_{4,1\bar10}(\Bb) = \Dt S_{2,01\bar1}(C_3\Bb),\n
		&\Dt S_{4,z}(\Bb) = \Dt S_{2,x}(C_3\Bb).
	\end{align}
	The conditions give,
	\begin{align}
		\Dt S_{1\perp}(C_3\Bb) =& g_1(\Bb)\hat x,\n
		\Dt S_{2\perp}(C_3\Bb) =& g_1(-\Bb)\hat x,\n
		\Dt S_{3\perp}(C_3\Bb) =& f_2^{even}(\Bb) \hat e_{01\bar1} + g_2^{even}(\Bb) \hat x,\n
		\Dt S_{4\perp}(C_3\Bb) =& -f_2^{even}(\Bb) \hat e_{01\bar1} + g_2^{even}(\Bb) \hat x,
	\end{align}
	so $\Dt S_{tot,\perp} (C_3\Bb) = g^{even}(\Bb) \hat x$. The physical meaning of this is just the rotation of $\Bb$ and $\Mb_{\perp}$.

	\section{Cluster Magnetic Multipoles in crystals\label{Ssec2}}
	
	The cluster magnetic multipoles (CMMs) are the quantification of a generic magnetic ordering, just like the magnetization~\cite{suzuki2017cluster}. In a magnetic unit cell, a spin cluster is defined as the group of atoms connected by the symmorphic symmetries. Therefore, there can be several spin clusters in a magnetic unit cell. The spin clusters are connected each other by the translational or non-symmorphic symmetries. 
	
	In a spin cluster $a$, one can define a CMM with rank $p$,
	\begin{align}
		M_{pq}^a =& \sqrt\f{4\pi}{2p+1}\sum_{i=1}^{N_a} \nabla(r_i^p Y_{pq})\cdot \mathbf{m}_i, \n
	\end{align}
	where $N_a$ is the number of atoms in the spin cluster, $r_i$ is the position of $i$-th atom, $Y_{pq}$ is the rank-$p$ spherical harmonics, and $\mathbf{m}_i$ is the magnetic moment at $i$-th atom. 
	After we calculate $M_{pq}$, we classified them into the bases of irreducible representations (IRREPs) of $T_d$, $O_h$, and $D_{3h}$ point group. 
	We are aware of the cluster magnetic toroidal multipoles, but we consider the response to the magnetic field only, so we do not consider this.
	We adopt the magnetic unit cell of pyrochlore oxides, Mn$_3$Ir, and CsMnBr$_3$ for each point group, whose structures are in the manuscript. 
	
\begin{figure*}[hbtp]
			\centering
			\includegraphics[width=2\columnwidth]{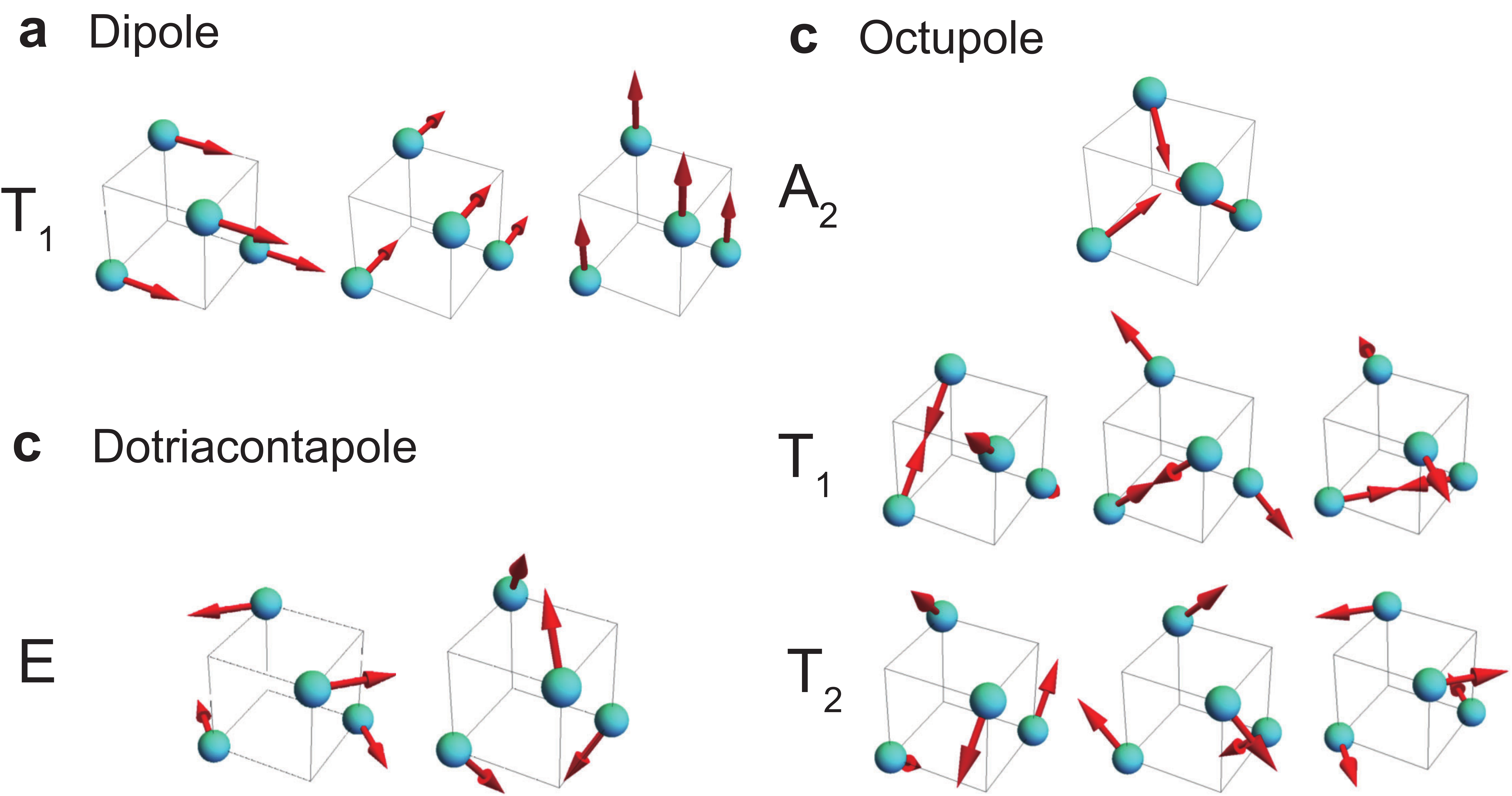}
			\caption{CMMs of pyrochlore oxides. (a) Dipoles, (b) octupoles, and (c) dotriacontapoles.}
			\label{SFig1}
		\end{figure*}

\begin{figure*}[h]
			\centering
			\includegraphics[width=2\columnwidth]{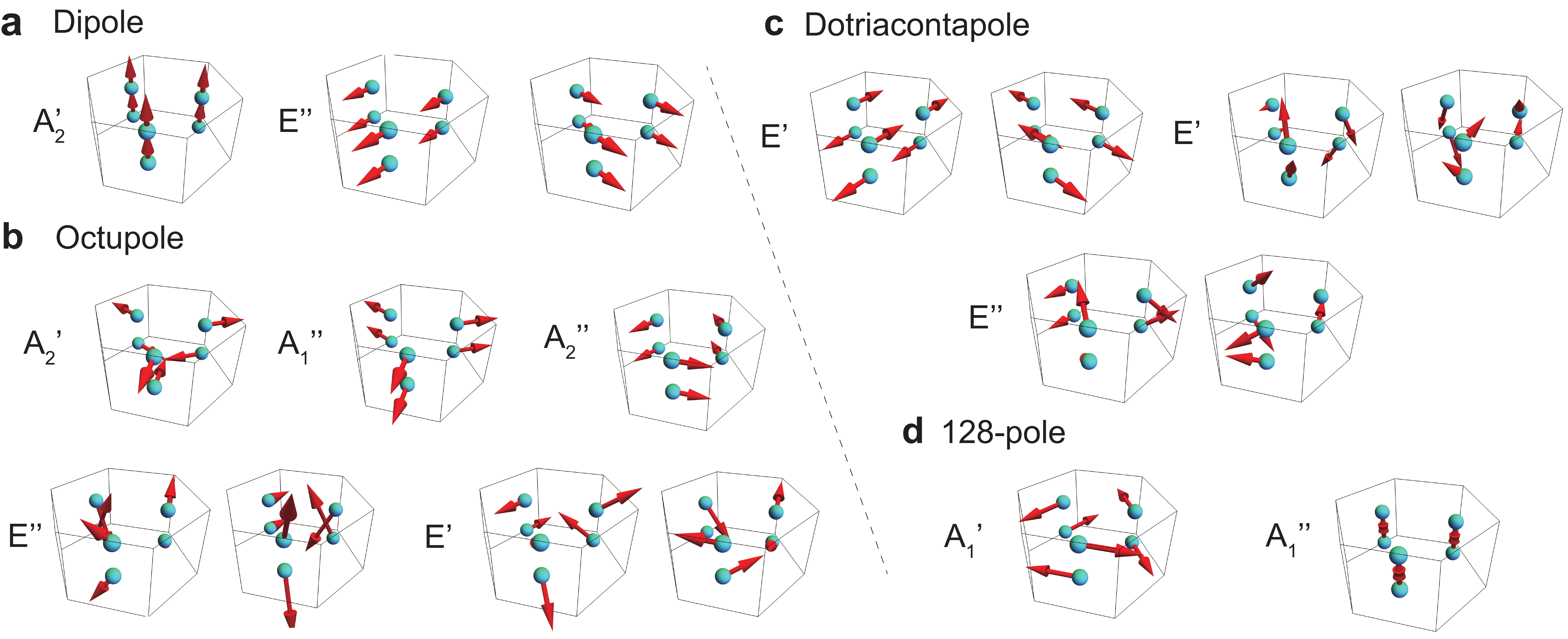}
			\caption{CMMs of CsMnBr$_3$. (a) Dipoles, (b) octupoles, (c) dotriacontapoles, and (d) 128-poles.}
			\label{SFig2}
		\end{figure*}

\begin{figure*}[h]
		\centering
		\includegraphics[width=1.5\columnwidth]{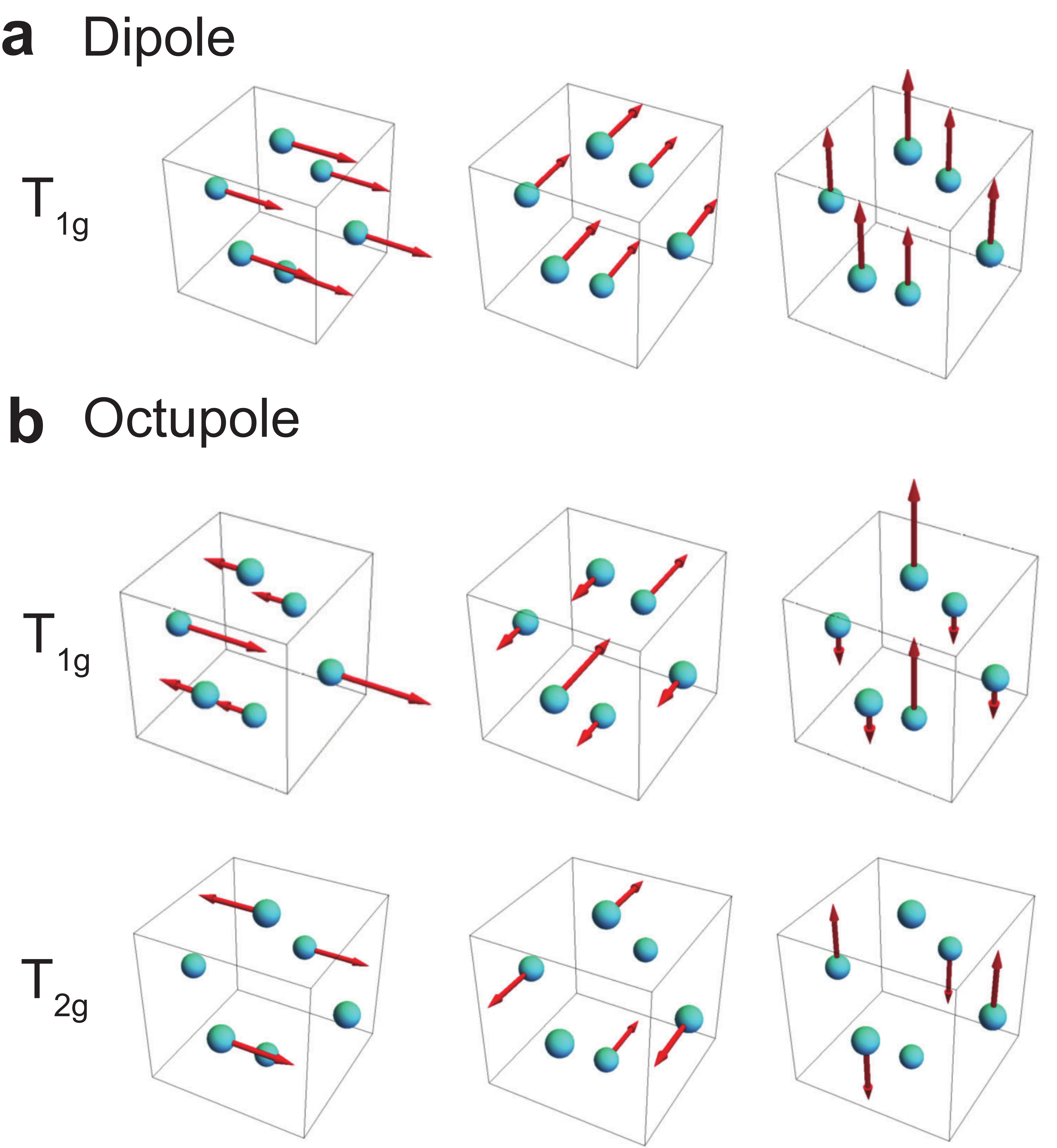}
		\caption{CMMs of Mn$_3$Ir. (a) Dipoles and (b) octupoles.}
		\label{SFig3}
	\end{figure*}

	Each magnetic unit cell of pyrochlore oxides, Mn$_3$Ir, and CsMnBr$_3$ is composed of a single spin cluster. 
	Note that the number of degrees of freedom in a spin cluster is $12$ ($4$ sublattices with $3$ axes) in pyrochlore oxides, $18$ ($6$ sublattices with $3$ axes) in CsMnBr$_3$, $9$ ($3$ sublattices with $3$ axes) in Mn$_3$Ir. Pyrochlore oxides can carry magnetic dipoles, octupoles, and dotriacontapoles. 
	Mn$_3$Ir carries magnetic dipoles and octupoles. 
	CsMnBr$_3$ carries magnetic dipoles, octupoles, dotriacontapoles, and 128-poles.
	Furthermore, we calculate the magnetic point group for each CMM, and determine whether TM exists or not, as we explain in the manuscript. 
	All results are in the Tables~\ref{TS1}-\ref{TS3}, and Figs.~\ref{SFig1}-\ref{SFig3}. 
	

		\begin{table*}[h]
			\centering
			\begin{tabular}{ c| c |c c c c c c c c c c c c c }
				\hline\hline
				Rank & CMM & $[100]$ & $[010]$ & $[001]$ & $[110]$ & $[1\bar10]$ & $[101]$ & $[\bar101]$ & $[011]$ & $[01\bar1]$ & $[111]$ & $[1\bar1\bar1]$ & $[\bar11\bar1]$ & $[\bar1\bar11]$ \\
				\hline
				Octupole & $A_2$ & $ \times $ & $ \times $ & $ \times $ & O & O & O & O & O & O & $ \times $ & $ \times $ & $ \times $ & $ \times $ \\
				&$T_{x}^1$ & $ \times $ & O & O & O & O & O & O & O & O & O & O & O & O \\
				&$T_{y}^1$ & O & $ \times $ & O & O & O & O & O & O & O & O & O & O & O \\
				&$T_{z}^1$ & O & O & $ \times $ & O & O & O & O & O & O & O & O & O & O \\
				&$T_{x}^2$ & $ \times $ & O & O & O & O & O & O & $ \times $ & $ \times $ & O & O & O & O \\
				&$T_{y}^2$ & O & $ \times $ & O & O & O & $ \times $ & $ \times $ & O & O & O & O & O & O \\
				&$T_{z}^2$ & O & O & $ \times $ & $ \times $ & $ \times $ & O & O & O & O & O & O & O & O \\ \hline
				Dotriacontapole & $E_1$ & $ \times $ & $ \times $ & $ \times $ & O & O & O & O & O & O & O & O & O & O \\
				&$E_2$ & $ \times $ & $ \times $ & $ \times $ & $ \times $ & $ \times $ & O & O & O & O & O & O & O & O \\
				\hline\hline
			\end{tabular}
			\caption{The presence (O) or absence (X) of TM for all possible magnetic structures in the pyrochlore lattice with the point group $T_d$ under various field directions.}
			\label{TS1}
		\end{table*}

		\begin{table*}[h]
			\centering
			\begin{tabular}{ c| c |c c c c c c c c c c c c c }
				\hline\hline
				Rank & CMM & $[100]$ & $[010]$ & $[001]$ & $[110]$ & $[1\bar10]$ & $[101]$ & $[\bar101]$ & $[011]$ & $[01\bar1]$ & $[111]$ & $[1\bar1\bar1]$ & $[\bar11\bar1]$ & $[\bar1\bar11]$ \\
				\hline
				Octupole & $A_1'$ & O & $ \times$ & $ \times $ & O & O & O & O & O & O & O & O & O & O \\
				& $A_2'$ & O & O & $ \times $ & O & O & O & O & O & O & O & O & O & O \\
				& $A_2''$ & $\times$ & O & $\times$ & O & O & O & O & O & O & O & O & O & O \\
				& $E_x'$ & $ \times $ & $ \times $ & $ \times $ & O & O & O & O & O & O & O & O & O & O \\
				& $E_y'$ & O & O & $ \times $ & O & O & O & O & O & O & O & O & O & O \\
				& $E_x''$ & $ \times $ & O & O & O & O & O & O & O & O & O & O & O & O \\
				& $E_y''$ & O & $ \times $ & O & O & O & O & O & O & O & O & O & O & O \\ \hline
				Dotriacontapole 
				& 2$E_x'$ & $\times$ & $\times$ & $\times$ & O & O & O & O & O & O & O & O & O & O \\
				& 2$E_y'$ & O & O & $\times$ & O & O & O & O & O & O & O & O & O & O \\
				& $E_x''$ & $ \times $ & O & O & O & O & O & O & O & O & O & O & O & O \\
				& $E_y''$ & O & $ \times $ & O & O & O & O & O & O & O & O & O & O & O \\ \hline
				128-pole & $A_1'$ & $\times$ & $\times$& $\times$ & O & O & O & O & O & O & O & O & O & O \\
				& $A_1''$ & O & $\times$ & $\times$ & O & O & O & O & O & O & O & O & O & O \\
				\hline\hline
			\end{tabular}
			\caption{The presence (O) or absence (X) of TM for all possible magnetic structures in CsMnBr$_3$ with the point group $D_{3d}$ under various field directions. The ground state is $A_1'$-128-pole.}
			\label{TS2}
		\end{table*}

		\begin{table*}[h]
			\centering
			\begin{tabular}{ c | c |c c c c c c c c c c c c c }
				\hline\hline
				Rank & CMM & $[100]$ & $[010]$ & $[001]$ & $[110]$ & $[1\bar10]$ & $[101]$ & $[\bar101]$ & $[011]$ & $[01\bar1]$ & $[111]$ & $[1\bar1\bar1]$ & $[\bar11\bar1]$ & $[\bar1\bar11]$ \\
				\hline
				Octupole & $T_{x}^1$ & $ \times $ & O & O & O & O & O & O & O & O & O & O & O & O \\
				& $T_{y}^1$ & O & $ \times $ & O & O & O & O & O & O & O & O & O & O & O \\
				& $T_{z}^1$ & O & O & $ \times $ & O & O & O & O & O & O & O & O & O & O \\ \hline 
				& $T_{x}^2$ & $ \times $ & O & O & O & O & O & O & $ \times $ & $ \times $ & O & O & O & O \\
				& $T_{y}^2$ & O & $ \times $ & O & O & O & $ \times $ & $ \times $ & O & O & O & O & O & O \\
				& $T_{z}^2$ & O & O & $ \times $ & $ \times $ & $ \times $ & O & O & O & O & O & O & O & O \\ \hline\hline
				Ground state & $T_{x}^1=T_y^1=T_z^1$ & O & O & O & O & O & O & O & O & O & $ \times $ & O & O & O
				\\
				\hline\hline
			\end{tabular}
			\caption{The presence (O) or absence (X) of TM for all possible magnetic structures in Mn$_3$Ir with the point group $O_h$ under various field directions.}
			\label{TS3}
		\end{table*}

	\clearpage

	\section{The numerical computations of $\Mb_{\perp}$\label{Ssec3}}
	
	\begin{figure*}[ht]
		\centering
		\includegraphics[width=2\columnwidth]{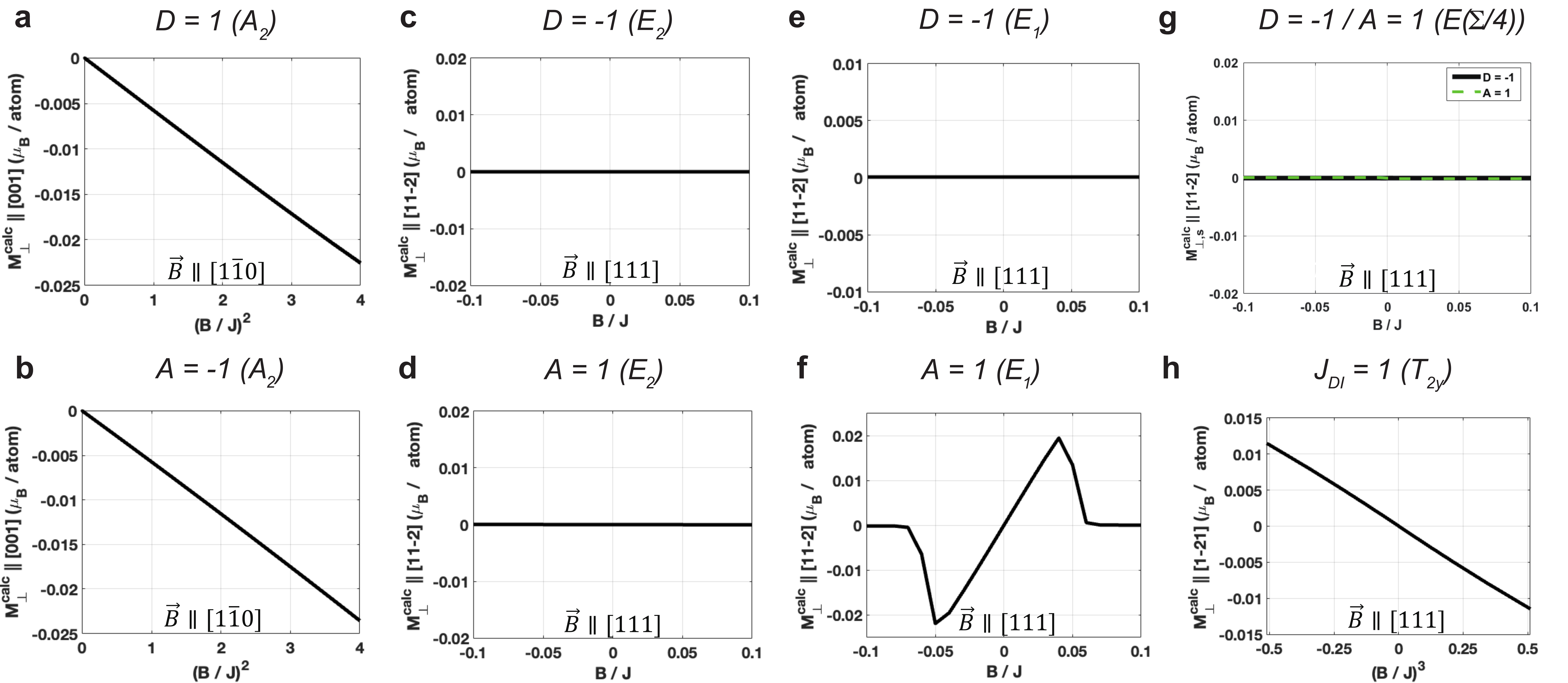}
		\caption{
			Numerical computation of $\Mb_{\perp}^{calc}$ from spin Hamiltonian. 
			(a-b) $\Mb_{\perp}^{calc}$ for $A_2$-order with (a) $D = 1$ and (b) $A = -1$ under $\Bb\parallel[1\bar10]$. 
			$\Mb_{\perp}^{calc}\propto B^2$ for both cases. 
			(c-d) $\Mb_{\perp}^{calc}$ for $E_2$-order with (c) $D = -1$ and (d) $A = 1$ under $\Bb \parallel[111]$. 
			$\Mb_{\perp}^{calc} = 0$ for both cases.
			(e-f) $\Mb_{\perp}^{calc}$ for $E_1$-order with (e) $D = -1$ and (f) $A = 1$ under $\Bb \parallel[111]$.
			In (e), $\Mb_{\perp}^{calc} = 0$, whereas in (f), $\Mb_{\perp}^{calc} \propto B$.
			(g) $\Mb_{\perp}^{calc}$ for $E(\pi/4)$-order with $D = -1$ (black solid line) and $A=1$ (black circles). $\Mb_{\perp}^{calc} = 0$.
			(h) $\Mb_{\perp}^{calc}$ for $T_{2y}$-order with $J_{DI} = 1$. $\Mb_{\perp}^{calc}\propto B^3$.
		}
		\label{SFig4}
	\end{figure*}
	
	Here, we perform numerical computations of $\Mb_{\perp}$ considering general Hamiltonians for classical spins on the pyrochlore lattice given by
	\begin{align}
		H=H_J+ H_{ani} - \sum_a \Bb \cdot \Sb_a, \label{eqHspin}
	\end{align}
	where $H_J$ indicate the isotropic Heisenberg spin Hamiltonian, and $H_{ani} = H_{DMI}$, $H_{SIA}$, $H_{DI}$ indicates the anisotropic spin Hamiltonian including
	Dzyaloshinskii-Moriya interaction (DMI)~\cite{elhajal2005ordering}, single-ion anisotropy (SIA)~\cite{glazkov2005single}, 
	and dipolar interaction (DI)~\cite{palmer2000order}, respectively. Their explicit forms are
	\begin{align}
		H_{J} =& J \sum_{\langle ab\rangle}  \Sb_a \cdot \Sb_b, \label{eqHs}\\
		H_{DMI} =& D \sum_{\langle ab\rangle} \hat \Db_{ab}\cdot (\Sb_a \times \Sb_b), \label{eqDM}\\
		H_{SIA} =& A\sum_a (\Sb_a\cdot\mathbf{n}_a)^2,\label{eqSIAs} \\
		H_{DI}=&J_{DI}\sum_{\langle ab\rangle} [\Sb_a\cdot \Sb_b - 3(\Sb_a\cdot \mathbf{r}_{ab})(\Sb_b\cdot \mathbf{r}_{ab})],\label{eqDIs}
	\end{align}
	where $J>0$ is the antiferromagnetic Heisenberg exchange interaction, $D$ $(A)$ is the strength of DMI (SIA), and $J_{DI}$ indicates the strength of the DI.
	Also, $\hat \Db_{ab}$ is the unit vector of DMI, $\mathbf{n}_a$ is the unit vector along the local $z$-axis of the spin $\Sb_a$, and $\mathbf{r}_{ab}$ is the normalized displacement vector between $a$-th and $b$-th sites.
	
	When we calculate the ground state, we begin from a random spin configuration and make an iterative approach. For given spin configuration and Hamiltonian $H$, the effective field at site $a$ is given by $\mathbf{h}_a=-\f{\pa H}{\pa \Sb_a}$. Then, we make the spin evolves to the direction of its mean-field,
	\begin{align}
		\Sb_a \rightarrow \f{\Sb_a + c\mathbf{h}_a}{||\Sb_a+c\mathbf{h}_a||}
	\end{align}
	where $c$ is the parameter given by hand. The ground state has the lowest energy among the converged spin configuration~\cite{sim2018discovery}.
	
	Depending on the type of the anisotropic spin interactions, various AFM ground states with distinct CMMs can appear~\cite{elhajal2005ordering,glazkov2005single,palmer2000order}.
	For example, suppose that $H = H_{J} + H_{DMI}$.
	Then, for $D>0$, the ground state is an $A_2$-octupole, 
	while for $D<0$, the ground state is an $E$-dotriacontapole or a $T_1$-octupole.
	On the other hand, if $H = H_{J} + H_{SIA}$, for $A<0$, the ground state is $A_2$-octupole, 
	while for $A>0$, the ground state is an $E$-dotriacontapole or a $T_2$-octupole.

	For each spin Hamiltonian on the pyrochlore lattice, we compute  $\Mb_{\perp} \equiv  \f{1}{4}\sum_{i=1}^{4} \langle\Sb_{i,\perp}\rangle$ as a function of $\Bb$. 
	The parameters are chosen as $J=|D|=|A|=1$ for numerical computations~\cite{sim2018discovery}.
	
	Let us first consider the $A_2$-order of $H_J + H_{DMI}$ and $H_J + H_{SIA}$.
	The relevant $\Mb_{\perp}\parallel\hat{z}$ under $\Bb \parallel [110]$ is plotted in Figs.~\ref{SFig4}a-b, respectively.
	One can clearly observe $\Mb_{\perp}\propto B^2$ for both cases.
	
	Next we consider $E=(E_1,E_2)$-order of $H_J+H_{DMI}$ and $H_J + H_{SIA}$.
	Here we choose $E_2$-order ($E=(0,1)$) as an initial ground state configuration, compute $\Mb_{\perp}\parallel[11\bar2]$ under $\Bb \parallel [111]$, and show the result in Figs.~\ref{SFig4}c-d.
	For both DMI and SIA cases, $\Mb_{\perp} = 0$.
	On the other hand, we choose $E_1$-order ($E=(1,0)$) as a ground state, compute $\Mb_{\perp}^{calc}\parallel[11\bar2]$ under $\Bb \parallel [111]$, and show the result in Fig.~\ref{SFig4}e-f.
	For DMI case, $\Mb_{\perp}^{calc} = 0$, but for SIA case, $\Mb_{\perp}^{calc} \propto B$.
	We choose $E(\pi/4)=(1/\sqrt2,1/\sqrt2)$-order as a ground state, compute $\Mb_{\perp}\parallel [11\bar2]$ under $\Bb \parallel [111]$, and show the result in Fig.~\ref{SFig4}g. For both $H_J + H_{DMI}$ and $H_J + H_{SIA}$, $\Mb_{\perp}^{calc} = 0$. In generic $E(\alpha)$-order ($E=(\sin\alpha, \cos\alpha)$), TM vanishes for both cases.
	
	Lastly, we consider $T_{2y}$-order of $H_J+H_{DI}$. Here we choose $T_{2y}$-order as a ground state. The resulting $\Mb_{\perp}^{calc}\parallel[1\bar21]$ is in Fig.~\ref{SFig4}h. One can easily note that $\Mb_{\perp}^{calc} \propto B^3$.
	
	\section{Hubbard model for pyrochlore oxides\label{Ssec4}}
	
	The Hubbard model for pyrochlore oxides is,
	\begin{align}
		H = H_0 + H_U^{MF} + H_B,
	\end{align}
	where
	\begin{align}
		H_0 = &\sum_{\langle ij\rangle} c_i^\da(t_1+it_2 \V d_{ij}\cdot \V \ma)c_j \n
		& + \sum_{\langle\langle ij \rangle\rangle} c_i^\da (t_1'+i(t_2'\V R_{ij} + t_3'\V D_{ij}))c_j,
	\end{align}
	is the kinetic Hamiltonian,
	\begin{align}
		H_U^{MF} = -U\sum_{i} (\langle \V j_i \rangle \cdot \V j_i - \langle \V j_i \rangle^2),
	\end{align}
	is the mean-field approximated Hubbard repulsion, and
	\begin{align}
		H_B = -\sum_i \V B \cdot \V j_i,
	\end{align}
	is the Zeeman coupling~\cite{witczak2013pyrochlore,oh2018magnetic}. Note that $\V j_i = \f{1}{2}\sum_i c_i^\da \V \ma c_j$. 
	
	The parameter we used are
	\begin{align}
		t_1 =& \f{130}{243}t_{oxy} + \f{17}{324}t_\ma - \f{79}{243}t_{\pi}, \n
		t_2 =& \f{28}{243}t_{oxy} + \f{15}{243}t_\ma - \f{40}{243}t_{\pi}, \n
		t_1' =& \f{233}{2916}t'_{\ma} - \f{407}{2187}t'_\pi, \n
		t_2' =& \f{1}{1458}t'_\ma + \f{220}{2187}t'_\pi ,\n
		t_3' =& \f{25}{1458}t_\ma'+\f{460}{2187}t_\pi' ,
	\end{align}
	where $t_{oxy}$ is the oxygen mediated nearest neighbor (NN) hopping, $t_\ma, t_{\pi}$ are the direct-overlap NN hopping, and $t_\ma',t_{\pi}'$ are the direct-overlap next NN hopping. We set $t_{oxy}=1$, $t_\ma = -0.8t_{oxy}$, $t_{\pi}=-2t_\ma/3$, and $t_{\ma,\pi}'=0.08t_{\ma,\pi}$. The Hubbard interaction $U=1.485$.
	Also, the Dzyaloshinskii-Moriya (DM) vectors are defined as
	\begin{align}
		\V d_{ij} =&~2\V a_{ij} \times \V b_{ij}, \n
		\V R_{ij} =&~\V b_{ik} \times \V b_{ki}, \n
		\V D_{ij} =&~\V d_{ik} \times \V d_{kj}.
	\end{align}
	where $\V b_{ij}$ is the vector from $i$-th to $j$-th sublattices, $\V a_{ij}$ is the vector that points from the center of a unit cell to the middle of $\langle ij\rangle $ bond, and $k$-th sublattice is the shared neighbor of $i$-th and $j$-th sublattices. We assume that the electric field is applied along $[1\bar10]$ and the magnetic field is applied in the $[111]$ plane and at angle $\ta$ by $[1\bar10]$. We fix $B/t_{oxy}=0.03$, changing $\ta=0$ to $2\pi$. We calculate the self-consistent ground state by using $30\times30\times30$ $k$-point mesh.
	
	Using the acquired self-consistent ground state, we compute the AHC and PHC. The formula of AHC and PHC are given by~\cite{nagaosa2010anomalous,nandy2017chiral} 
	\begin{align}
		\ma_{xy}^{AHE} =& \f{e^2}{\hbar} \sum_n \int \f{d^3k}{(2\pi)^3} \Om_n^z(\kb) f(\ep_n(\kb)), \label{eqAHC}\\
		\ma_{xy}^{PHE} =& e^2 \int \f{d^3k}{(2\pi)^3} D\tau(-\f{\pa f}{\pa \ep})(v_x+\f{eB_x}{\hbar}(\vb_k\cdot\Omb_k)) \n &\times (v_y + \f{eB_y}{\hbar} (\vb_k\cdot\Omb_k)), \label{eqPHC}
	\end{align}
	where $\Omb_k$ is the Berry curvature, $\vb_k$ is the group velocity, $f(\ep)$ is the Fermi-Dirac distribution, $\tau$ is the relaxation time, $D=(1+(e/\hbar)\Bb\cdot\Omb_k)^{-1}$ is the phase volume change because of Berry curvature. We take $e=\hbar=a=\tau=1$, where $a$ is the lattice constant.

	\section{Details in microscopic origin of TM\label{Ssec5}}
	
	\subsection{Without anisotropy}
	
	When there is no anisotropy, the spin interaction is given by
	$H=H_J$ with $J>0$. The ground state of $H_J$ is highly degenerate. When we consider the pyrochlore lattice, the Hamiltonian is 
	\begin{align}
		H_J = J\sum_{\langle ab \rangle } \Sb_a \cdot \Sb_b = 8  J N_c\Mb^2 - \f{J N_c}{2}\sum_{a=1}^4 \Sb_a^2,
	\end{align}
	where $N_c$ is the number of unit cell, $\Mb = \f{1}{4}\sum_{a=1}^4 \Sb_a$ is the average magnetization. Physically, the classical spins have the same magnitude, $\Sb_a^2=1$ for all $a$s, so the last term is a constant as $\sum_{a=1}^4 \Sb_a^2=4$. Then,
	\begin{align}
		H_J = 8JN_c \Mb^2.
	\end{align}
	The ground state can be any antiferromagnetic states, i.e. $\Mb = 0$.
	
	When we add the magnetic field, the Hamiltonian is
	\begin{align}
		H =& H_J + H_B \n
		=& 8 J N_c \Mb^2 - 4 N_c \Bb \cdot \Mb - \f{J N_c}{2}\sum_{a=1}^4 \Sb_a^2,
	\end{align}
	Again, the last term is constant, the Hamiltonian becomes
	\begin{align}
		H = 8J N_c (\Mb-\f{\Bb}{4J})^2.
	\end{align}
	The energy minimum is $H=0$ at
	\begin{align}
		\Mb = \f{\Bb}{4J}.
	\end{align}
	Since the magnetization is parallel to $\Bb$, the TM vanishes.

	\subsection{Single-ion anisotropy $(A>0)$}
	
	Next, let us add the SIA,
	\begin{align}
		H= &H_J + H_{SIA} + H_B,\\
		=& J \sum_{\langle ab\rangle}  \Sb_a \cdot \Sb_b+A\sum_a (\Sb_a\cdot\mathbf{n}_a)^2- \sum_a \Bb \cdot \Sb_a, \label{eqSIAv}
	\end{align}
	where $J>0$ and $A>0$. For $H_{SIA}$, the energy minimum requires
	\begin{align}
		\Sb_a \cdot \mathbf{n}_a = 0, \label{eqSIAcond}
	\end{align}
	for each $a = 1,2,3,4$, separately. 
	This forces each spin to lying in its local $xy$-plane, so reduces the ground state manifold. 
	The red arrows are spins $\Sb_a$, the yellow arrows are the hard axes $\mathbf{n}_a$, and the yellow planes are the local-$xy$ plane.
	Thus, for the full Hamiltonian $H=H_J + H_{SIA} + H_B$, the energy minimum requires
	\begin{align}
		\Mb = \f{\Bb}{4J},\quad \Sb_a\cdot \mathbf{n}_a = 0. \label{eqSIA2}
	\end{align}
	When $\Bb=0$, the ground state is either $E$-dotriacontapole or $T_2$-octupole, where all spins are lying on their local $xy$-plane.
	$E$-dotriacontapole has $E_1$ and $E_2$-orders. 
	There are several ground states. The first ground state is generally represented by $E(\alpha) = (E_1=\sin\alpha,E_2=\cos\alpha)$-order. The spins in $E(\alpha)$-order are
	\begin{align}
		\Sb_a = \cos\alpha \hat x_a + \sin\alpha\hat y_a \label{eqEa}
	\end{align}
	where
	\begin{align}
		&\hat x_1 = [1\bar10], \hat x_2 = [110], \hat x_3 = [\bar1\bar10], \hat x_4 = [\bar110], \n
		&\hat y_1 = [11\bar2], \hat y_2 = [1\bar12], \hat y_3 = [\bar112], \hat y_4 = [\bar1\bar1\bar2],\n
		&\hat z_1 = [111], \hat z_2 = [1\bar1\bar1], \hat z_3 = [\bar11\bar1], \hat z_4 = [\bar1\bar11]. \label{eqlocal}
	\end{align}
	Note that $\hat x_a$ is the direction of spins in $E_2$-order, and $\hat y_a$ is that in $E_1$-order.
	
	The next type of ground state is $E+T_2(\A)$-orders. There are three kinds of the orders. The first one is $E_1+T_{2z}(\A)$ which is represented by
	\begin{align}
		\Sb_a = d_a \cos\A \hat x_a + \sin\A \hat y_a
	\end{align}
	where $d_a = 1$ for $a=1,4$ and $d_a = -1$ for $a=2,3$. 
	The spin configuration for $\alpha = 0$ is $T_{2z}$-order while that for $\A = \pi/2$ is $E_1$-order. 
	This means that when $\A$ increases two spins at $a=2,3$ in $E_1+T_{2z}(\A)$-order rotate oppositely from those spins in $E(\A)$-order. 
	
	The next one is $E(\pi/6)+T_{2x}(\A)$ which is represented by
	\begin{align}
		\Sb_a = e_a \cos\A \hat x_a' + \sin\A \hat y_a' 
	\end{align}
	where $\hat x_a' = \f{1}{2}\hat x_a -\f{\sqrt3}{2}\hat y_a, \hat y_a'= \f{\sqrt3}{2}\hat x_a + \f{1}{2}\hat y_a$, $e_a = 1$ for $a=1,2$ and $e_a = -1$ for $a=3,4$.  The spin configuration for $\alpha = 0$ is $T_{2x}$-order while that for $\A = \pi/2$ is $E(\pi/6)$-order. This means that when $\A$ increases two spins at $a=3,4$ in $E(\pi/6)+T_{2x}(\A)$-order rotate oppositely from those spins in $E(\A)$-order. 
	
	The last one is $E(5\pi/6)+T_{2y}(\A)$ which is represented by
	\begin{align}
		\Sb_a = f_a \cos\A \hat x_a'' + \sin\A \hat y_a'' 
	\end{align}
	where $\hat x_a'' = \f{1}{2}\hat x_a +\f{\sqrt3}{2}\hat y_a, \hat y_a''= -\f{\sqrt3}{2}\hat x_a + \f{1}{2}y_a$, $f_a = 1$ for $a=1,3$ and $f_a = -1$ for $a=2,4$. The spin configuration for $\alpha = 0$ is $T_{2y}$-order while that for $\A = \pi/2$ is $E(5\pi/6)$-order. This means that when $\A$ increases two spins at $a=2,4$ in $E(5\pi/6)+T_{2y}(\A)$-order rotate oppositely from those spins in $E(\A)$-order.
	
	Please note that all ground state manifold is continuously degenerate without magnetic field. We take $E(\A)$-order as a ground state for convenience, but the following results are the same as the other ground state manifolds as well.
	
	When $\Bb\neq0$, the stationary condition satisfying Eq.~\ref{eqSIA2} usually exists, but not in a few cases. Because all spins are described by their polar and azimuthal angles ($\ta_a,\phi_a$), the number of degrees of freedom is $8$. As the number of equations in Eq.~\ref{eqSIA2} is $7$, the ground state exists in general. However, as we seek the stationary condition by a smooth deviation of ($\ta_a$,$\phi_a$) at small $\Bb$ limit, the such state satisfying Eq.~\ref{eqSIA2} may not exist depending on the initial spin configuration: for instance, $E(\pi/6 + n\pi/3)$-orders. ($n\in \mathbb{Z}$)
	
	For illustration, let us consider the case when the initial spin configuration is a $E$-dotriacontapole $E=(E_1, E_2)$ at $\Bb=0$. 
	In the case of $E_2$-order with $\alpha = 0$,
	the configuration of each spin can be represented by
	\begin{align}
		\Sb_a =& \sin(\f\pi 2+\ta_a)\cos(\phi_a)\hat x_a 
		\n&+ \sin(\f\pi 2+\ta_a)\sin(\phi_a)\hat y_a \n& + \cos(\f\pi 2+\ta_a)\hat z_a, 
	\end{align}
	where $\hat x_a, \hat y_a, \hat z_a$ are in Eq.~\ref{eqlocal}.
	When $\ta_a,\phi_a = 0$, the spin configuration is the $E_2$-order.
	To describe the deformation of the spin configuration under small $\Bb$, we expand $\Sb_a$ up to the first order of angular variables as
	\begin{align}
		\Sb_a = \hat x_a + \phi_a \hat y_a - \ta_a \hat z_a, \label{eqE2}
	\end{align}
	with which Eq.~\ref{eqSIA2} become
	\begin{align}
		\Sb_a& \cdot \mathbf{n}_a = -\ta_a = 0. \n
		\Mb =& \f{1}{4\sqrt3}[\phi_1+\phi_2-\phi_3-\phi_4-\ta_1-\ta_2+\ta_3+\ta_4,\n &\phi_1-\phi_2+\phi_3-\phi_4-\ta_1+\ta_2-\ta_3+\ta_4,\n&2(-\phi_1+\phi_2+\phi_3-\phi_4)-\ta_1+\ta_2+\ta_3-\ta_4]\n
		=&\f{\Bb}{4J}.   \label{eqSIAsol1}   
	\end{align}
	The solution of the first constraint equations is $\theta_a=0$, which means that the SIA forces the spins to rotate in their local xy-planes. 
	For $\Bb=[B_x,B_y,B_z]$, the solution of the second constriant equations can be written as
	\begin{align}
		\phi_2=& (-2\sqrt3 B_y+\sqrt3 B_z)/4J+\phi_1,\n \phi_3=& (-2\sqrt3 B_x+\sqrt3 B_z)/4J+\phi_1,\n \phi_4 =& - \sqrt3(B_x+B_y)/2J+\phi_1. \label{eqsol1}
	\end{align} 
	That is, the stationary condition satisfying Eq.~\ref{eqSIA2} exists. Accordingly, the TM should vanish. We confirm that for a general $E(\alpha)$-order case, the stationary condition satisfying Eq.~\ref{eqSIA2} exists and thus the TM vanishes.
	
	On the other hand, there are some cases that the stationary condition satisfying Eq.~\ref{eqSIA2} does not exist. Considering the $E_1$-order with $\alpha = \pi/2$, 
	the configuration of each spin can be represented by
	\begin{align}
		\Sb_a =& \sin(\f\pi 2+\ta_a)\cos(\f\pi 2+\phi_a)\hat x_a 
		\n&+ \sin(\f\pi 2+\ta_a)\sin(\f\pi 2+\phi_a)\hat y_a \n& + \cos(\f\pi 2+\ta_a)\hat z_a. \label{eqspin}
	\end{align}
	Up to the first order of angular variables, we find
	\begin{align}
		\Sb_a = -\phi_a \hat x_a + \hat y_a - \ta_a \hat z_a, \label{eqE12}
	\end{align}
	with which Eq.~\ref{eqSIA2} becomes
	\begin{align}
		\Sb_a \cdot \mathbf{n}_a =& - \ta_a = 0. \n
		\Mb = \f{1}{6}[& (3\sqrt2(-\phi_1-\phi_2+\phi_3+\phi_4)\n&+2\sqrt3 (-\ta_1-\ta_2+\ta_3+\ta_4)),\n
		&(3\sqrt2(\phi_1-\phi_2+\phi_3-\phi_4)\n&+2\sqrt3 (-\ta_1+\ta_2-\ta_3+\ta_4)),\n&
		\f{-\ta_1+\ta_2+\ta_3-\ta_4}{\sqrt3}]=\f{\Bb}{4J}.  \label{eqSIA3}
	\end{align}
	Again, the SIA forces the condition $\theta_a=0$. However, the second equations of $\phi_a$ does not have a solution when $B_z\neq0$. 
	The stationary condition satisfying Eq.~\ref{eqSIA2} does not exist.
	Considering the threefold and twofold rotations in $H_J + H_{SIA}$, $\alpha = \pi/6 + n\pi$ does not have a solution when $B_x \neq 0$, $\alpha = 5\pi/6 + n\pi$ does not have a solution when $B_y \neq 0$, and $\alpha = \pi/2 + n\pi$ does not have a solution when $B_z \neq 0.$

	For an arbitrary $E(\alpha)$-order, the spin configuration is given by
	\begin{align}
		\Sb_a =& \cos\ta_a\cos(\alpha+\phi_a)\hat x_a + \cos\ta_a \sin (\alpha+\phi_a)\hat y_a \n&- \sin\ta_a \hat z_a,
	\end{align}
	Then, Eq.~\ref{eqSIA2} gives
	\begin{align}
		\Sb_a\cdot \mathbf{n}_a =& - \ta_a = 0, \n
		\Mb =& \f{\Bb}{4J} \n
		=& \f{1}{4} [\f{1}{6}(\sqrt6 \cos\A-3\sqrt2 \sin\A)(\phi_1 +\phi_2 -\phi_3 - \phi_4),\n
		&~\f{1}{6}(\sqrt6 \cos\A + 3\sqrt2 \sin\A) (\phi_1 -\phi_2 +\phi_3 - \phi_4),\n
		&~-\sqrt\f{2}{3}(\phi_1-\phi_2-\phi_3+\phi_4)\cos\A  ]. \label{eqgeneralE}
	\end{align}
	When $\Bb = B (b_x,b_y,b_z)$, the solution is given by $\ta_a = 0$ and 
	\begin{align}
		\phi_2 =& \phi_1-\f{\sqrt3B(2\sqrt6b_y-\sqrt6 b_z - 3\sqrt2 b_z \tan\A)}{4J \cos\A (\sqrt3+3\tan\A)},\n
		\phi_3 =& \phi_1-\f{\sqrt3B(2\sqrt6b_x-\sqrt6 b_z + 3\sqrt2 b_z \tan\A)}{4J \cos\A (\sqrt3-3\tan\A)},\n
		\phi_4 =& \phi_1-\f{3B(\sqrt6(b_x+b_y)+3\sqrt2(b_x-b_y)\tan\A)}{2J \cos\A (\sqrt3+3\tan\A)(\sqrt3-3\tan\A)},
	\end{align}
	
	where $\phi_1$ is arbitrary. The solution does not exist when $\A = \pi/6+n\pi/3$ where the denominator goes to 0.

	In fact, there is a condition that $\A = \pi/6 + n\pi/3$ does not have a solution. For example, $\A = \pi/2 + n\pi$, the second equation of Eq.~\ref{eqgeneralE} becomes
	\begin{align}
		\f{\Bb}{J}=&\f{3\sqrt2}{6} [(\mp (\phi_1 +\phi_2 -\phi_3 - \phi_4),\pm  (\phi_1 -\phi_2 +\phi_3 - \phi_4),\n
		&~0 ],
	\end{align}
	Therefore, whenever $B_z \neq 0$, the solution does not exist. 
	Similarly, for $\A = \pi/6+n\pi$ ($5\pi/6+n\pi$), the solution does not exist whenever $B_x \neq 0$ ($B_y\neq 0$).

	$E(\pi/6+n\pi/3)$-orders are special since all spins in the orders are either parallel or antiparallel to the projected $\Bb$ onto local-$xy$ plane simultaneously. For $E(\alpha \neq \pi/6 + n\pi/3)$-orders, there are some spins not parallel to the projected $\Bb$. For example, in $E(\pi/3)$-order,
	\begin{align}
		\Sb_1 \parallel [10\bar1], \Sb_2 \parallel [101], \Sb_3 \parallel [\bar101], \Sb_4 \parallel [\bar10\bar1]. \label{eqEpi3}
	\end{align}
	Suppose that
	\begin{align}
		\Bb = B(\cos\beta [10\bar1]/\sqrt2 + \sin\beta [111]/\sqrt3),
	\end{align}
	then $\Sb_1$ is parallel to the projected $\Bb$ onto its local-$xy$ plane whenever $\beta \neq \pi/2 + n\pi$. However, for other spins, the projected $\Bb$ onto spin $a$ ($\Bb^P_a$) are
	\begin{align}
		\Bb^{P}_2 =& \cos\beta [101]/\sqrt3 + (2\sqrt3 \cos\beta + 3\sqrt2 \sin\beta) [12\bar1]/18,\n
		\Bb^P_3 =& -\sin\beta [\bar101]/\sqrt3 + (\cos\beta) [121]/3\sqrt3,\n
		\Bb^P_4 =& -\cos\beta [\bar10\bar1]/\sqrt3 + (-2\sqrt3 \cos\beta + 3\sqrt2 \sin\beta) [1\bar21]/18,
	\end{align}
	These are not parallel to $E(\pi/3)$-order in Eq.~\ref{eqEpi3} generally.
	
	\begin{figure}[t]
		\centering
		\includegraphics[width=\columnwidth]{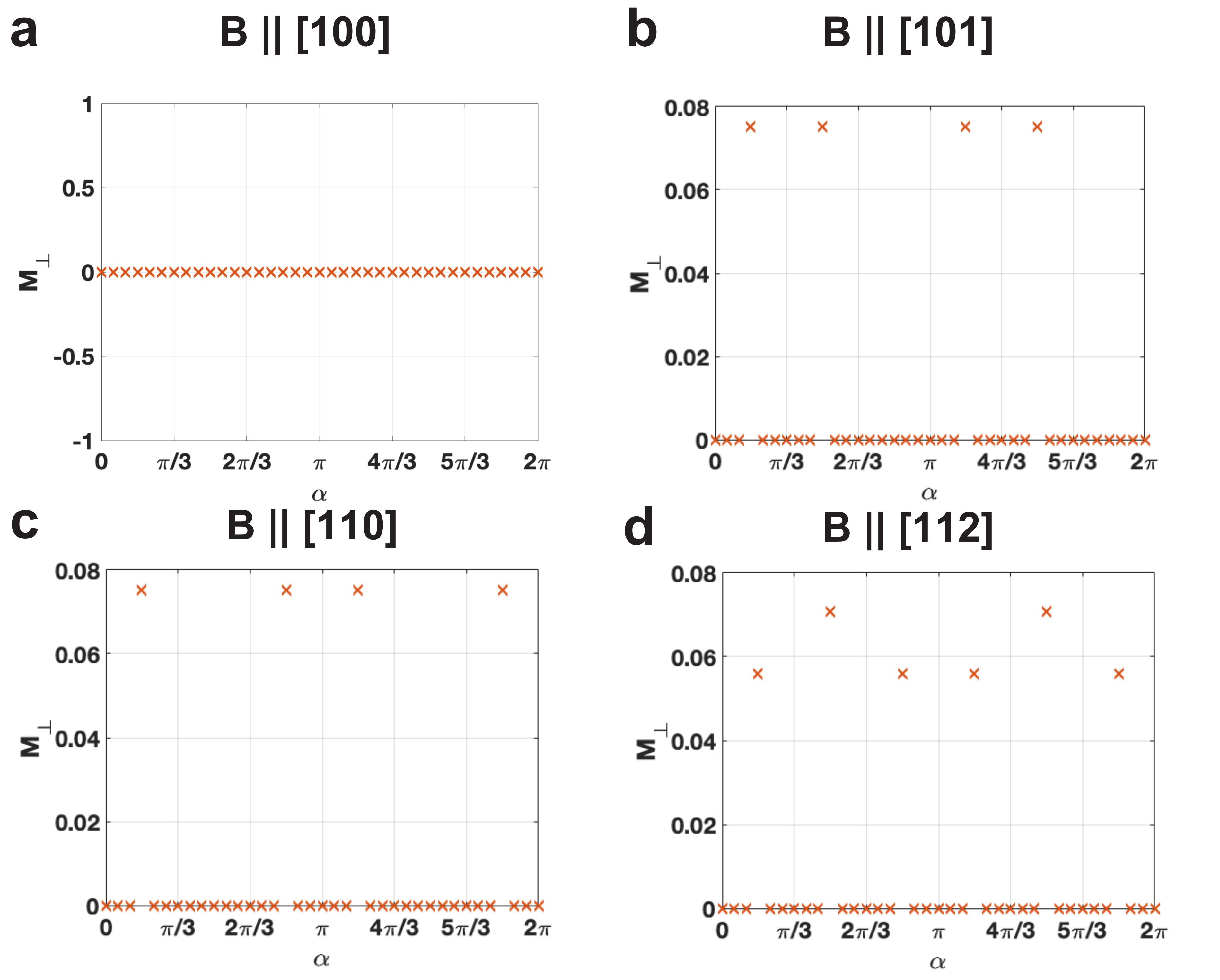}
		\caption{TM $\Mb_{\perp}$ for $E(\alpha)$-order with SIA under (a) $\Bb \parallel [100]$, (b) $\Bb \parallel [101]$, (c) $\Bb \parallel [110]$ and (d) $\Bb \parallel [112]$.}
		\label{SFig5}
	\end{figure}
	
	However, when $\Bb \parallel [001]$, the projected $\Bb$ on each local-$xy$ plane is 
	\begin{align}
		\Bb^P_1 \parallel [\bar1\bar12], \Bb^P_2 \parallel [1\bar12], \Bb^P_3 \parallel [\bar112], \Bb^P_4 \parallel [112]
	\end{align}
	All projected magnetic fields are either parallel or antiparallel to $E_1$-order ($E(\pi/2), E(3\pi/2)$). Hence, whenever $B_z \neq 0$, all spins deviate from local-$xy$ planes. Similarly, for $\Bb \parallel [100]$, all projected magnetic fields are along $E(\pi/6)$ or $E(7\pi/6)$, and for $\Bb \parallel [010]$, all projected magnetic fields are along $E(5\pi/6)$ or $E(11\pi/6)$. Hence, all spins deviate from local-$xy$ planes whenever $B_x\neq 0$ for $E(\pi/6)$ and $E(7\pi/6)$, and whenever $B_y\neq 0$ for $E(5\pi/6)$ and $E(11\pi/6)$. 
	
	We can obtain $(\ta_a,\phi_a)$ of the stationary condition in series of $B = |\Bb|$ in $E_1$-order by the derivative of Hamiltonian,
	\begin{align}
		\f{\pa H}{\pa \ta_a} = 0,\quad \f{\pa H}{\pa \phi_a} = 0. \label{eqHder}
	\end{align}
	Up to the first order of $B$,
	\begin{align}
		&\ta_1 = \ta_4 = -\ta_2 = -\ta_3  = \f{B}{6A+4J}, \n
		&\phi_1 = \phi_4 = 0,  \n
		&\phi_2 = -\phi_3 = -\f{B}{\sqrt6 J}, \label{eqE1}
	\end{align}
	All spins deviate from local-$xy$ planes since $\ta_a \neq 0$. The TM of the stationary condition is
	\begin{align}
		\Mb_{\perp}^{\text{SIA}} =& \f{\sqrt2 }{4(2+3A/J)}(\f{A}{J})(\f{B}{J})\hat e_{11\bar2}. \label{eqspinPM1}
	\end{align}
	This is consistent with the numerical results in Fig.~\ref{SFig4}.

	We present some parts of results by rotating $\Bb$ in the manuscript.
	We also try other cases in Fig~\ref{SFig5}. 
	When $\Bb \parallel [100]$, $\Mb_{\perp}$ vanishes for all $\alpha$ because of the twofold rotation symmetry.
	When $\Bb \parallel [101]$, $\Mb_{\perp}$ is finite only for $\alpha = \pi/6+n\pi$ and $\pi/2+n\pi$.
	When $\Bb \parallel [110]$, $\Mb_{\perp}$ is finite only for $\alpha = \pi/6+n\pi$ and $5\pi/6+n\pi$.
	When $\Bb \parallel [112]$,  $\Mb_{\perp}$ is finite only for $\alpha = \pi/6 + n\pi/3$.
	
	\subsection{Single-ion anisotropy $(A<0)$}
	
	Let us consider Eq.~\ref{eqSIAv} again, but $A<0$. Then, since $\mathbf{n}_a$ is an easy axis, the spins point to the easy axis. In pyrochlore lattice, the ground state of $H_J + H_{SIA}$ is an $A_2$-order. 
	The ground state manifold is now discretely degenerate, so the TM usually emerges when the symmetry admits.
	
	When $\Bb$ is applied to $A_2$-order, each spin is confined in the plane spanned by its easy axis and $\Bb$. Let us try $\Bb \parallel [001]$ first, where the twofold rotation symmetry makes the TM vanishes. $A_2$-order can be represented by Eq.~\ref{eqspin}, but the local axes are
	\begin{align}
		&\hat x_1 = [\bar1\bar12], \hat x_2 = [1\bar12], \hat x_3 = [\bar112], \hat x_4 = [112],\n
		&\hat y_1 = [111], \hat y_2 = [1\bar1\bar1], \hat y_3 = [\bar11\bar1], \hat y_4 = [\bar1\bar11],\n
		&\hat z_1 = [\bar110], \hat z_2 = [110], \hat z_3 = [\bar1\bar10], \hat z_4 = [1\bar10].
	\end{align}
	Note that $x_ay_a$-plane is spanned by $A_2$-order and $\Bb$. Without $\Bb$, $\ta_a = \phi_a = 0$ for all $a$. The stationary condition can be found by Eq.~\ref{eqHder} up to the second order of B.
	\begin{align}
		&\ta_a = 0,
		\phi_1 = \phi_4 = - \sqrt\f{3}{2}\f{B}{4J+3A}  + \f{3}{2\sqrt2}\f{B^2}{(4J+3A)^2},\n
		&	\phi_2 = \phi_3 = - \sqrt\f{3}{2}\f{B}{4J+3A}  - \f{3}{2\sqrt2}\f{B^2}{(4J+3A)^2}. \label{eqSIAX}
	\end{align}
	Note that for any order of $B$, $\ta_a = 0$. The spins are confined within local $x_ay_a$-plane. The magnetization is $\Mb = \f{\Bb}{4J+3A}$.
	
	This is different from the energy minimum condition. By adding some constant to Eq.~\ref{eqSIAv}, we have
	\begin{align}
		H =& -\f{4}{3}A[ \sum_{r=1}^3 (\sqrt2 M_r - \sum_a \Sb_a \cdot \mathbf{P}_a^r)^2 \n&+ 3((\sum_a \Sb_a \cdot \hat x_a)^2 + (\sum_a \Sb_a \cdot \hat y_a)^2 \n&+ \sum_{r=1}^3(\sum_a \Sb_a \cdot \mathbf{T}_a^r)^2)] + 8J (\Mb - \f{\Bb}{4J})^2,
	\end{align}
	where $\hat x_a$ and $\hat y_a$ are defined in Eq.~\ref{eqlocal},
	\begin{align}
		&	\mathbf{P}_1^1 = [011], \mathbf{P}_2^1 = [0\bar1\bar1], \mathbf{P}_3^1 = [0\bar11], \mathbf{P}_4^1 = [01\bar1], \n
		&	\mathbf{P}_1^2 = [101], \mathbf{P}_2^2 = [\bar101], \mathbf{P}_3^2 = [\bar10\bar1], \mathbf{P}_4^2 = [10\bar1], \n
		&	\mathbf{P}_1^3 = [110], \mathbf{P}_2^3 = [\bar110], \mathbf{P}_3^3 = [1\bar10], \mathbf{P}_4^3 = [\bar1\bar10], \label{eqT1}
	\end{align}
	and
	\begin{align}
		&	\mathbf{T}_1^1 = [01\bar1], \mathbf{T}_2^1 = [0\bar11], \mathbf{T}_3^1 = [0\bar1\bar1], \mathbf{T}_4^1 = [011], \n
		&	\mathbf{T}_1^2 = [\bar101], \mathbf{T}_2^2 = [101], \mathbf{T}_3^2 = [10\bar1], \mathbf{T}_4^2 = [\bar10\bar1], \n
		&	\mathbf{T}_1^3 = [1\bar10], \mathbf{T}_2^3 = [\bar1\bar10], \mathbf{T}_3^3 = [110], \mathbf{T}_4^3 = [\bar110].\label{eqT2}
	\end{align}
	Note that $\Mb = (M_1,M_2,M_3) = (M_x,M_y,M_z)$. $\mathbf{P}_a^r$ and $\mathbf{T}_a^r$ is related to $T_1$ and $T_2$-octupole, respectively.
	Considering that the constants are positive, the energy minimum conditions are
	\begin{align}
		&	\Mb = \f{\Bb}{4J}, \n
		&	\sum_a \Sb_a \cdot \hat x_a = 0, \n
		&	\sum_a \Sb_a \cdot \hat y_a = 0,	\n
		&	\sum_a \Sb_a \cdot \mathbf{T}_a^r = 0~(r=1,2,3),\n
		&	\sum_a \Sb_a \cdot \mathbf{P}_a^r = \sqrt2 M_r~ (r=1,2,3).
	\end{align}
	The magnetization of the energy minimum condition is $\Mb=\Bb/4J$ which is different from stationary condition. This is because the stationary condition satisfying the energy minimum conditions generally does not exist since we have a total of 8 variables $(\ta_a,\phi_a)$ and a total of 11 equations.
	
	Let us try $\Bb\parallel [101]$, where the symmetries admits the TM. $A_2$-order can be represented by Eq.~\ref{eqspin}, but the local axes are
	\begin{align}
		&\hat x_1 = [1\bar21], \hat x_2 = [101], \hat x_3 = [121], \hat x_4 = [101],\n
		&\hat y_1 = [111], \hat y_2 = [1\bar1\bar1], \hat y_3 = [\bar11\bar1], \hat y_4 = [\bar1\bar11],\n
		&\hat z_1 = [\bar101], \hat z_2 = [12\bar1], \hat z_3 = [\bar101], \hat z_4 = [1\bar2\bar1].
	\end{align}
	Note that local-$x_ay_a$ plane is again spanned by $A_2$-order and $\Bb$.
	Without $\Bb$, $\ta_a=\phi_a =0$. The stationary condition is obtained by Eq.~\ref{eqHder}. Up to the second order of $B$,
	\begin{align}
		&	\ta_1=\ta_3 =0, \ta_2 = -\ta_4 = \f{9J B^2}{2\sqrt2 (3A+4J)^3}.\n
		&	\phi_1 = -\f{\sqrt3B}{2(3A+4J)}  + \f{3\sqrt2(3A+J)B^2}{4(3A+4J)^3}, \n
		&	\phi_2 = \phi_4 = -\f{3B}{(3A+4J)} \n
		&	\phi_3 = -\f{\sqrt3B}{2(3A+4J)}  - \f{3\sqrt2(3A+J)B^2}{4(3A+4J)^3}, 
	\end{align}
	The TM is given by
	\begin{align}
		\Mb_{\perp}^{SIA,A_2}= \f{9\sqrt3 }{8(3A/J+4)^3} (\f{A}{J})(\f{B^2}{J^2})\hat y.
	\end{align}
	
\begin{figure}[t]
		\centering
		\includegraphics[width=\columnwidth]{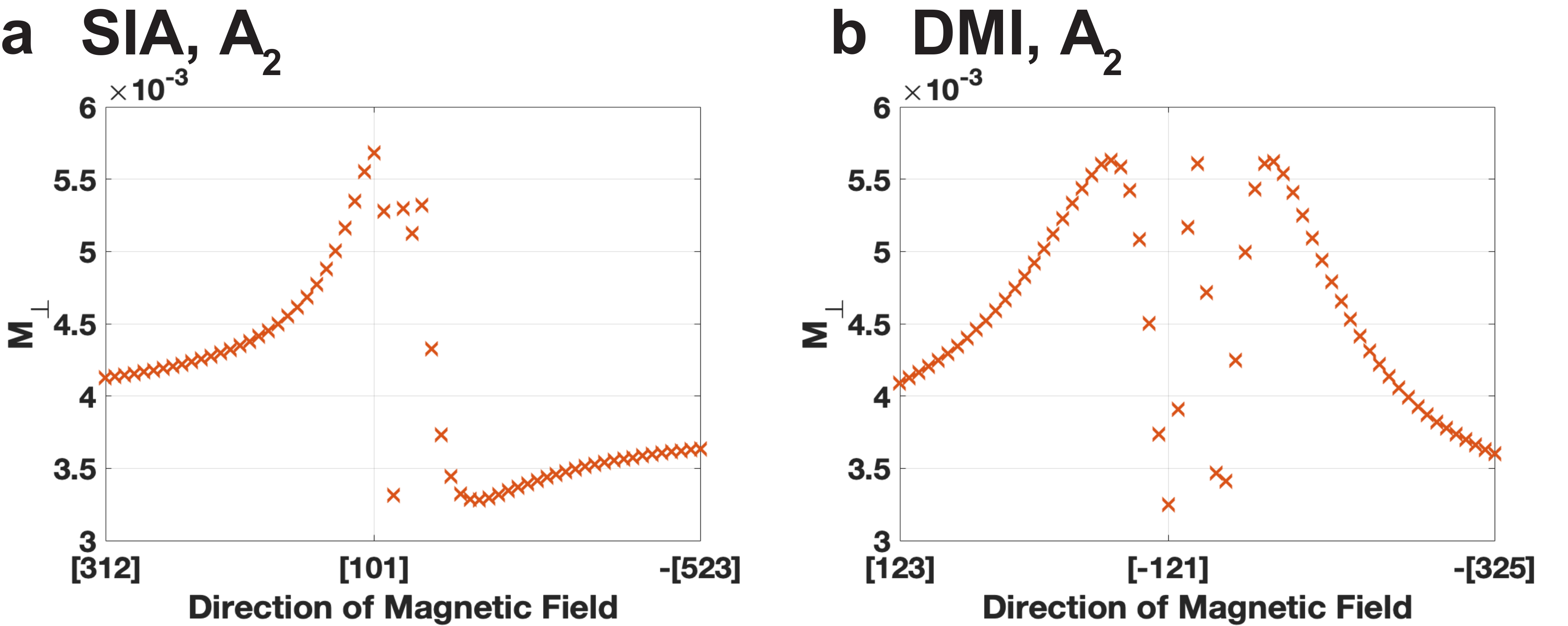}
		\caption{The TM $\Mb_{\perp}$ for $A_2$-order, (a) with SIA under $\Bb$ changing from $[312]$ to $[101]$ to $[\bar5\bar2\bar3]$, and (b) with DMI under $\Bb$ changing from $[123]$ to $[\bar121]$ to $[\bar3\bar2\bar5]$.}
		\label{SFig6}
	\end{figure}

	We also find TM under $\Bb$ in an arbitrary direction. In Fig.~\ref{SFig6}a, we present the change of TM under rotating $\Bb$ from $[312]$ to $[101]$, and to $[\bar5\bar2\bar3]$ in sequence. The TM arises for every direction.
	
	\subsection{Dzyaloshinskii-Moriya interaction ($D > 0$)}
	
	Next, let us consider DMI $(D>0)$ and $A_2$-order. 
	\begin{align}
		H =& H_J + H_{DMI} + H_B \n
		=& J\sum_{\langle ab\rangle} \Sb_a\cdot\Sb_b + D \sum_{\langle ab \rangle} \hat \Db_{ab}\cdot (\Sb_a\times \Sb_b) - \Bb \cdot \sum_a \Sb_a. \label{eq50}
	\end{align}
	
	The role of DMI is to confine two spins in the plane perpendicular to the DM vector $\Db_{ij}$ because the energy is minimized when $\Sb_i \times \Sb_j$ is anti-parallel to $\Db_{ij}$. 
	When we consider the unit cell of pyrochlore lattice, each spin prefers to be perpendicular to the surrounding six DM vectors. The perpendicular direction to the DM vectors is the local-$z$ axis of $\Sb_i$. 
	Hence, DMI confines $\Sb_i$ to its local-$z$ axis, and the ground state is $A_2$-octupole, as same as the previous section. 
	Again, the ground state manifold is now discretely degenerate, and the TM can arise when the symmetry admits.
	
	When $\Bb$ is applied, it is natural that each spin is confined in the plane spanned by its local-$z$ axis and $\Bb$. We try other directions from the previous section since the result is similar.
	When $\Bb \parallel [100]$, the TM is canceled by twofold rotation symmetry. $A_2$-order can be represented by Eq.~\ref{eqspin}, but the local axes are changed by
	\begin{align}
		&\hat x_1 = [2\bar1\bar1], \hat x_2 = [211], \hat x_3 = [21\bar1], \hat x_4 = [2\bar11], \n
		& \hat y_1 = [111], \hat y_2 = [1\bar1\bar1], \hat y_3 = [\bar11\bar1], \hat y_4 = [\bar1\bar11], \n
		&\hat z_1 = [0\bar11], \hat z_2=[01\bar1], \hat z_3 = [011], \hat z_4 = [0\bar1\bar1].
	\end{align}
	Note that $x_a y_a$-plane is spanned by $A_2$-order and $\Bb$. Without $\Bb$, $\ta_a = \phi_a = 0$ for all $a$. The stationary condition under $B$ can be found by Eq.~\ref{eqHder} up to the second order of $B$. 
	\begin{align}
		&\ta_a = 0,  \n
		&\phi_1 = \phi_2 = -\sqrt{\f32}\f{B}{7D+4J}+\f{3B^2}{2\sqrt2(7D+4J)^2}, \n
		&\phi_3 = \phi_4 = -\sqrt{\f32}\f{B}{7D+4J}-\f{3B^2}{2\sqrt2(7D+4J)^2}.
	\end{align}
	Note that for any order of $B$, $\ta_a = 0$. Hence, the spins are confined within $x_a y_a$-plane. The magnetization is just $\Mb = \f{\Bb }{7D+4J}$. 
	
	Please note that this is different from the energy minimum condition of DMI. By adding some constant to Equation~\ref{eq50}, we have 
	\begin{align}
		H =& 8(J+D)(\Mb-\f{\Bb}{4(J+D)})^2 + 4D[\n&3(\sum_a \Sb_a \cdot \hat x_a/4)^2 + 3(\sum_a \Sb_a \cdot \hat y_a/4)^2 \n&+ 3 \sum_{r=1}^3 (\sum_a \Sb_a \cdot \mathbf{P}_a^r/4)^2 \n& + \sum_{r=1}^3 (\sum_a \Sb_a \cdot \mathbf{T}_a^r/4)^2],
	\end{align}
	where $\hat x_a$ and $\hat y_a$ are defined in Eq.~\ref{eqlocal}, and $\mathbf{P}_a^r$ and $\mathbf{T}_a^r$ are defined in Eq.~\ref{eqT1} and \ref{eqT2}.
	Considering that all coefficients are positive, the energy minimum condition is given by
	\begin{align}
		&	\Mb = \f{\Bb}{4(J+D)}, \n
		&	\sum_a \Sb_a \cdot x_a = 0, \n
		&	\sum_a \Sb_a \cdot y_a = 0, \n
		&	\sum_a \Sb_a \cdot \mathbf{P}_a^r = 0 ~(r=1,2,3), \n
		&	\sum_a \Sb_a \cdot \mathbf{T}_a^r = 0 ~(r=1,2,3),	
	\end{align}
	Here, the magnetization of energy minimum condition is $\Mb = \f{\Bb}{4(J+D)}$ which is different from stationary condition. This is because the stationary condition satisfying the energy minimum conditions generally does not exist for discretely degenerate case, as we have a total of 8 variables $(\ta_a,\phi_a)$ and a total of 11 equations. 
	
	For $\Bb \parallel [110]$, on the other hand, the TM is admitted by symmetry breaking. $A_2$-order can be represented again by Eq.~\ref{eqspin}, but the local axes are now
	\begin{align}
		&		\hat x_1 = \hat x_4 = [1\bar10], \hat x_2 = [\bar11\bar2], \hat x_3 = [\bar112] \n
		&		\hat y_1 = [111], \hat y_2 = [1\bar1\bar1], \hat y_3 = [\bar11\bar1], \hat y_4 = [\bar1\bar11], \n
		&		\hat z_1 = [11\bar2], \hat z_2 = \hat z_3 = [110], \hat z_4 = [112].
	\end{align}
	Note that $x_a y_a$-plane is spanned by $A_2$-order and $\Bb$.	Without $\Bb$, $\ta_a = 0$ and $\phi_a = 0$. The stationary condition under $B$ can be found by Eq.~\ref{eqHder}. Up to the second order of $B$
	\begin{align}
		&		\ta_1 = \ta_4 = 0, \ta_2 = -\ta_3 = \f{9(D-2J)}{4\sqrt2(7D+4J)^3}B^2, \n
		&		\phi_1 = - \f{\sqrt3}{2(7D+4J)}B + 3\f{(17D+2J)}{4\sqrt2(7D+4J)^3}B^2, \n
		&		\phi_2 = \phi_3 = -\f{3}{2(7D+4J)}B, \n
		&		\phi_4 = - \f{\sqrt3}{2(7D+4J)}B - 3\f{(17D+2J)}{4\sqrt2(7D+4J)^3}B^2.
	\end{align}
	Since $\ta_{2,3} \neq 0$, $\Sb_2$ and $\Sb_3$ deviate from the plane spanned by $\Bb$ and $\hat y_a$. The TM in the stationary condition is 
	\begin{align}
		\Mb_{\perp}^{DM} = \f{27\sqrt3}{8(4 + 7D/J)^3}(\f{D}{J})(\f{B^2}{J^2}) \hat z.
	\end{align}
	
	We also find TM under $\Bb$ in an arbitrary direction. In Fig.~\ref{SFig6}b, we present the change of TM under rotating $\Bb$ from $[123]$ to $[\bar121]$, and to $[\bar3\bar2\bar5]$ in sequence. The TM arises for every direction.
	
	\subsection{Dzyloshinskii-Moriya Interaction ($D <0$)}
	
	Next, let us consider DMI ($D<0$). $A_2$-order is not ground state anymore because $A_2$-order gains energy. 
	We can acquire the energy minimum condition by adding some constants to Eq.~\ref{eq50}. 
	\begin{align}
		H =& -12D (\sum_a \Sb_a \cdot \hat v_a/4)^2 - 8D \sum_{r=1}^3[(\sum_{a} \Sb_a \cdot \mathbf{T}_{a}^{r}/4)^2] \n&+8(J-D/2)(\Mb-\f{\Bb}{4(J-D/2)})^2,
	\end{align}
	where
	\begin{align}
		\hat v_1 = [111], \hat v_2 = [1\bar1\bar1], \hat v_3 = [\bar11\bar1], \hat v_4 = [\bar1\bar11],
	\end{align}
	and $\mathbf{T}_a^r$ is defined in Eq.~\ref{eqT2}.
	Since $D<0$, all coefficients are positive, so the energy minimum conditions are
	\begin{align}
		&\sum_a \Sb_a \cdot \hat v_a = 0, \n
		&\sum_a \Sb_a \cdot \mathbf{T}_a^r = 0~(r=1,2,3), \n
		&\Mb = \f{\Bb}{4(J-D/2)}. \label{eqDMcond}
	\end{align}
	If $\Bb = 0$, the energy minimum spin configuration can be found by setting 
	\begin{align}
		\Sb_a =& \cos\phi_a\sin\ta_a\hat x_a \n&+ \sin\phi_a\sin\ta_a\hat y_a +\cos\ta_a \hat z_a,
	\end{align} 
	whose local axes are defined in Eq.~\ref{eqlocal}. Then, the conditions give rise to $\ta_a = \pi/2$ and $\phi_1=\phi_2=\phi_3=\phi_4=\alpha$, which corresponds to $E(\alpha)$-order. 
	Furthermore, we find that $E + T_1$-orders are also the energy minimum spin configurations, which is represented by
	\begin{align}
		\Sb_a = \cos\beta \hat x_a + \sin\beta \hat y_a,
	\end{align}  
	where the local axes are
	\begin{align}
		&\hat x_1 = [011], \hat x_2 = [0\bar1\bar1], \hat x_3 = [0\bar11], \hat x_4 = [01\bar1], \n
		&\hat y_1 = [0\bar11], \hat y_2 = [01\bar1], \hat y_3 = [0\bar1\bar1], \hat y_4 = [011],\n
		& \hat z_a = [100],
	\end{align}
	or
	\begin{align}
		&\hat x_1 = [101], \hat x_2 = [\bar101], \hat x_3 = [\bar10\bar1], \hat x_4 = [10\bar1], \n
		&\hat y_1 = [10\bar1], \hat y_2 = [101], \hat y_3 = [\bar101], \hat y_4 = [\bar10\bar1],\n
		& \hat z_a = [010],
	\end{align}
	or
	\begin{align}
		&\hat x_1 = [110], \hat x_2 = [\bar110], \hat x_3 = [1\bar10], \hat x_4 = [\bar1\bar10], \n
		&\hat y_1 = [1\bar10], \hat y_2 = [110], \hat y_3 = [\bar1\bar10], \hat y_4 = [\bar110],\n
		& \hat z_a = [00\bar1].
	\end{align}
	Note that $\hat x_a$ are the spin directions in $T_{1x}, T_{1y}$, and $T_{1z}$-orders, $\hat y_a$ are that in $E(-\pi/3)$, $E(\pi/3)$, and $E_2$-orders in sequence. The energy of $E+T_1$-order is the same as $E(\alpha)$-order, because $\Sb_a \times \Sb_b$ remain invariant while $\B$ varies.
	Note that for each $E+T_1$-order, all spins are in the same plane. For example, all spins in $E_2+T_{1z}$-order are in $xy$-plane, those in $E(\pi/3)+T_{1y}$-order are in $xz$-plane, and those in $E(-\pi/3)+T_{1x}$-order are in $yz$-plane.
	
	Since the ground state is continuously degenerate, TM usually vanishes when $\Bb$ is applied. 
	Let us choose a general $E(\alpha)$-order as a ground state for convenience. For small magnetic field $\Bb$, the spins are described by
	\begin{align}
		\Sb_a =& \cos(\alpha+\phi_a)\cos(\ta_a)\hat x_a \n&+ \sin(\alpha+\phi_a)\cos(\ta_a)\hat y_a -\sin\ta_a \hat z_a
	\end{align}
	where the local axes are defined in Eq.~\ref{eqlocal}, and $(\ta_a,\phi_a)$ is the angle deviation by $\Bb$. Up to the first order of angles, the spins are expanded as
	\begin{align}
		\Sb_a =& (\cos\alpha-\phi_a \sin\alpha)\hat x_a \n&+ (\phi_a\cos\alpha+\sin\alpha)\hat y_a -\ta_a \hat z_a.
	\end{align}
	We put the expansion in Eq.~\ref{eqDMcond},
	\begin{align}
		&\sum_a \Sb_a \cdot \hat v_a = -\f{1}{4}\sum_a \ta_a =0, \n
		&\sum_a \Sb_a \cdot \mathbf{T}_a^1 = \f{1}{8}(\phi_1+\phi_2-\phi_3-\phi_4)\n &\times(\sqrt3 \cos\alpha + \sin\alpha)=0, \n
		&\sum_a \Sb_a \cdot \mathbf{T}_a^2 = -\f{1}{8}(\phi_1-\phi_2+\phi_3-\phi_4)\n &\times(\sqrt3 \cos\alpha - \sin\alpha)=0, \n
		&\sum_a \Sb_a \cdot \mathbf{T}_a^3 = \f{1}{4}(-\phi_1 + \phi_2 + \phi_3 - \phi_4)\sin\alpha=0,	\nonumber
	\end{align}
	and
	\begin{align}
		&\Mb =  [\f{1}{24}(-\phi_1-\phi_2+\phi_3+\phi_4)(\sqrt6 \cos\alpha - 3\sqrt2 \sin\alpha)  \n& + \f{1}{4\sqrt3}(-\ta_1-\ta_2+\ta_3+\ta_4) , \n
		&  \f{1}{24}(\phi_1-\phi_2+\phi_3-\phi_4)(\sqrt6 \cos\alpha + 3\sqrt2 \sin\alpha) \n& + \f{1}{4\sqrt3}(-\ta_1+\ta_2-\ta_3+\ta_4), \n
		& -\f{\sqrt2}{4\sqrt3} (\phi_1-\phi_2-\phi_3+\phi_4)\cos\alpha \n& + \f{1}{4\sqrt3}(-\ta_1+\ta_2+\ta_3-\ta_4)]=  \f{\Bb}{4(J-D/2)}.
	\end{align}
	
	We find the solution by expanding $(\ta_a,\phi_a)$ as a series of $B$ and and take terms only up to $B$. 
	When $B=|\Bb|$,
	\begin{align}
		\ta_a =& -3 \hat z_a \cdot \f{\Bb}{4(J-D/2)}, \phi_a = q \f{B}{4(J-D/2)}, \label{eqDMsol}
	\end{align}
	where $q$ is an arbitrary constant, and $\hat z_a$ is in Eq.~\ref{eqlocal}.
	Accordingly, we acquire the magnetization,
	\begin{align}
		\Mb = \f{\Bb}{4(J-D/2)}. \label{eqDMTM}
	\end{align}
	Please note that when we try Eq.~\ref{eqHder} to find the stationary condition, we have the same result in Eqs.~\ref{eqDMsol}-\ref{eqDMTM} for the general $E(\alpha)$-order. This is different from SIA $A>0$ case, where all spins are confined to the local-$xy$ planes.
	
	We also analytically find that the stationary condition also gives $\Mb = \f{\Bb}{(J-D/2)}$ for arbitrary $E(\alpha)$-order. 
	The TM vanishes for any $\alpha$. 
	This is consistent with numerical calculations in a generic $E(\alpha)$-order, as shown in Figs.~\ref{SFig4}c,e,g. 
	It is still valid that the TM generally vanishes in continuous degenerate ground states.
	
	We perform the same procedure to $E+T_1$-orders. For $E_2 + T_{1z}(\B)$-order, for example, the spins under $\Bb$ are described by
	\begin{align}
		\Sb_a =& \cos\ta_a\cos(\B+\phi_a)\hat x_a + \cos\ta_a\sin(\B+\phi_a)\hat y_a \n&+ \sin\ta_a \hat z_a,
	\end{align} 
	where
	\begin{align}
		& \hat x_1 = [1\bar10],	\hat x_2 = [110],	\hat x_3 = [\bar1\bar10],	\hat x_4 = [\bar110], \n
		& \hat y_1 = [110],	\hat y_2 = [\bar110],	\hat y_3 = [1\bar10],	\hat y_4 = [\bar1\bar10], \n
		& \hat z_a = [00\bar1].
	\end{align}
	The expansion gives rise to
	\begin{align}
		\Sb_a =& (\cos\B - \sin\B \phi_a) \hat x_a + (\sin\B + \cos\B \phi_a) \hat y_a \n&+ \ta_a \hat z_a
	\end{align}
	The energy minimum condition in Eq.~\ref{eqDMcond} gives
	\begin{align}
		&\sum_a \Sb_a \cdot \hat v_a = \f{1}{4\sqrt3} (\ta_1-\ta_2-\ta_3+\ta_4 \n&+ \sqrt2 \cos\B (\phi_1-\phi_2-\phi_3+\phi_4)) =0, \n
		&\sum_a \Sb_a \cdot \mathbf{T}_a^1 = \f{1}{4\sqrt2}((\phi_1-\phi_2+\phi_3-\phi_4)\cos\B \n&+ (-\phi_1-\phi_2+\phi_3+\phi_4)\sin\B)=0, \n
		&\sum_a \Sb_a \cdot \mathbf{T}_a^2 = \f{1}{4\sqrt2}((\phi_1+\phi_2-\phi_3-\phi_4)\cos\B \n&+ (\phi_1-\phi_2+\phi_3-\phi_4)\sin\B)=0, \n
		&\sum_a \Sb_a \cdot \mathbf{T}_a^3 = \f{1}{4}(\sum_a \ta_a)=0,	\nonumber
	\end{align}
	and
	\begin{align}
		&\Mb =  [\f{1}{8}( \sqrt2 (\ta_1+\ta_2-\ta_3-\ta_4) + (\phi_1-\phi_2+\phi_3-\phi_4)\cos\B  \n& + (\phi_1+\phi_2-\phi_3-\phi_4)\sin\B ) , \n
		&  \f{1}{8}( \sqrt2 (\ta_1+\ta_2-\ta_3-\ta_4) + (-\phi_1-\phi_2+\phi_3+\phi_4)\cos\B  \n& + (\phi_1-\phi_2+\phi_3-\phi_4)\sin\B ), \n
		& -\f{1}{4}(\phi_1-\phi_2-\phi_3+\phi_4)\sin\B]=  \f{\Bb}{4(J-D/2)}.
	\end{align}
	When $\Bb = B (b_x,b_y,b_z)$ and $B=|\Bb|$, the solution of the system of equations is
	\begin{align}
		\ta_1 =& \f{B}{4(J-D/2)}(b_z + (b_x+b_y) \cos(2\B) \n&+(-b_x+b_y) \sin(2\B)),\n
		\ta_2 =& \f{B}{4(J-D/2)}(b_z + (-b_x+b_y) \cos(2\B)\n& +(-b_x-b_y) \sin(2\B)),\n
		\ta_3 =& \f{B}{4(J-D/2)}(b_z + (b_x-b_y) \cos(2\B)\n& +(b_x+b_y) \sin(2\B)),\n
		\ta_4 =& \f{B}{4(J-D/2)}(b_z + (-b_x-b_y) \cos(2\B)\n& +(b_x-b_y) \sin(2\B)),\nonumber
	\end{align}
	\begin{align}
		\phi_2 =& \phi_1 -  \f{2\sqrt2B}{4(J-D/2)} (b_x \cos\B+b_y\sin\B),\n
		\phi_3 =& \phi_1 +  \f{2\sqrt2B}{4(J-D/2)} (-b_y \cos\B+b_x\sin\B),\n
		\phi_4 =& \phi_1 +  \f{2\sqrt2B}{4(J-D/2)} (-(b_x+b_y) \cos\B \n& +(b_x-b_y)\sin\B),
	\end{align}
	The TM vanishes for arbitrary $\B$ as well.

	\subsection{Dipolar interaction}
	
	\begin{figure}
		\centering
		\includegraphics[width=\columnwidth]{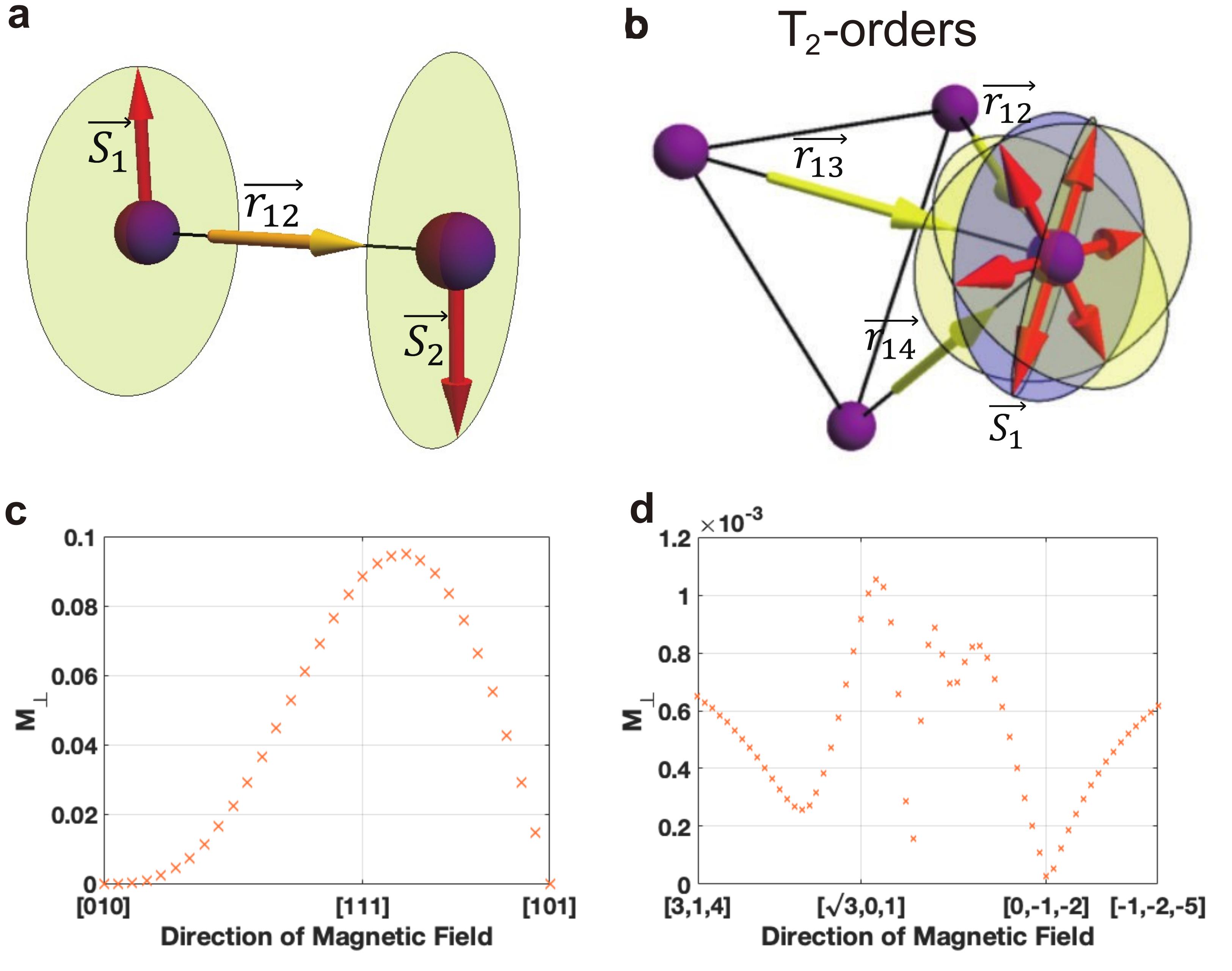}
		\caption{(a) The role of DI. The spins $\Sb_1$ and $\Sb_2$ that are apart by $\mathbf{r}_{12}$ are confined in the plane perpendicular to $\mathbf{r}_{12}$. (a-b) The yellow arrows are the displacement vectors, the red arrows are the spins, the yellow planes are the planes perpendicular to displacement vectors. (b) $\Sb_1$ is surrounded by $\mathbf{r}_{12}$, $\mathbf{r}_{13}$, $\mathbf{r}_{14}$. However, the intersecting line of planes perpendicular to $\mathbf{r}_{12}$, $\mathbf{r}_{13}$, and $\mathbf{r}_{14}$ (yellow planes) is absent. Instead, $\Sb_1$ is at the intersection of one of such three planes and the blue plane perpendicular to the $\mathbf{r}_{12}+\mathbf{r}_{13}+\mathbf{r}_{14} \propto [111]$. Accordingly, the ground state of pyrochlore lattice with DI is three $\mathbf{T}_2$-orders up to time-reversal. We choose $T_{2y}$-order as a ground state. (c) Changing $\Bb$ from $[010]$ to $[111]$ and to $[101]$, $\Mb_{\perp}$ for $T_{2y}$-order are plotted. Only at $\Bb \parallel [010]$ or $[101]$, $\Mb_{\perp}$ vanishes. (d) Changing $\Bb$ from $[314]$ to $[\sqrt301]$, to $[0\bar1\bar2]$, and to $[\bar1\bar2\bar5]$ in sequence, $\Mb_{\perp}$ are plotted.}
		\label{SFig7}
	\end{figure}

	Lastly, let us discuss DI ($J_{DI}>0$) and $T_{2y}$-order.
	\begin{align}
		H =& H_J + H_{DI} + H_B, \n
		=& J \sum_{\langle ab \rangle}\Sb_a\cdot \Sb_b + J_{DI} \sum_{\langle ab \rangle } [\Sb_a\cdot \Sb_b - 3(\Sb_a\cdot \mathbf{r}_{ab})(\Sb_b\cdot \mathbf{r}_{ab})] \n&- \Bb \cdot \sum_a \Sb_a.
	\end{align}
	To get an insight for DI, let us first consider a 2-spin system with $\Sb_1$ and $\Sb_2$ aparted by $\mathbf{r}_{12}$ interacting with $H=H_{J}+H_{DI}$, as shown in Fig.~\ref{SFig7}a. There is a competition between $H_J$ and $H_{DI}$, since $H_J$ prefers the antiferromagnet and $H_{DI}$ prefers the ferromagnet along $\mathbf{r}_{12}$. However, since $J$ is much stronger than $J_{DI}$, the ground state is an antiferromagnet. 
	Instead, $H_{DI}$ makes two spins confined in the plane perpendicular to $\mathbf{r}_{ij}$. 
	Two spins can freely rotate within the plane, but they are at the opposite direction to each other. 
	
	When we have the unit cell of pyrochlore lattice, there are three displacement vectors for each spin. 
	Note that we consider only the nearest neighbor DI for convenience~\cite{palmer2000order}. 
	For example, $\Sb_1$ is surrounded by $\mathbf{r}_{12} = [011]/\sqrt2$, $\mathbf{r}_{13} = [101]/\sqrt2$, and $\mathbf{r}_{14} = [110]/\sqrt2$ (see Fig.~\ref{SFig7}b). The planes perpendicular to $\mathbf{r}_{12}, \mathbf{r}_{13}$, and $\mathbf{r}_{14}$ have no intersecting lines. 
	Instead, the energy minimum is on one of three planes and perpendicular to $\mathbf{r}_{12}+\mathbf{r}_{13}+\mathbf{r}_{14}  \propto [111]$, which is indicated by red arrows in Fig.~\ref{SFig7}c. This argument are the same for the other spins. We have total 6 minimums for each spin as follows.
	\begin{align}
		&\Sb_1 : \pm [01\bar1], \pm [\bar101], \pm [1\bar10], \n
		&\Sb_2 : \pm [0\bar11], \pm [101], \pm [\bar110], \n
		&\Sb_3 : \pm [0\bar1\bar1], \pm [10\bar1], \pm [110],\n
		&\Sb_4 : \pm [011], \pm[\bar10\bar1], \pm [\bar110].
	\end{align}	
	
	When we choose one of 6 minimums of $\Sb_1$, the other spins are automatically chosen. For example, let us choose $\Sb_1 \parallel [\bar101]$. Since $H_J + H_{DI}$ prefers two spins pointing opposite directions, $\Sb_3 \parallel [10\bar1]$. The energy from interacting $\Sb_1,\Sb_3$ and $\Sb_2,\Sb_4$ is minimized when $\Sb_2 \parallel [101]$, $\Sb_4 \parallel [\bar10\bar1]$ because $\Sb_2, \Sb_4 \perp \Sb_1, \Sb_3$. Note that $\Sb_2$ and $\Sb_4$ are pointing opposite to each other. This spin configuration corresponds to $T_{2y}$-order. Other choices give $T_{2x}, T_{2z}$-orders, similarly. Unlike two site case, where ferromagnetic and antiferromagnetic orders compete, $T_{2i} (i=x,y,z)$-order is always the ground state for any $J_{DI}>0$ in the unit cell of pyrochlore lattice. The ground state is now discretely degenerate, and the TM usually arises when the symmetry admits.
	
	Let us find the ground state under $\Bb$. When $\Bb \parallel [010]$, the TM vanishes by twofold rotation symmetry. $\Sb_a$ is represented by Eq.~\ref{eqspin}, whose local axes are
	\begin{align}
		&\hat x_1 = [121], \hat x_2 = [\bar121], \hat x_3 = [\bar12\bar1], \hat x_4 = [12\bar1], \n
		&\hat y_1 = [\bar101], \hat y_2 = [101], \hat y_3 = [10\bar1], \hat y_4 = [\bar10\bar1], \n
		&\hat z_1 = [1\bar11], \hat z_2 = [11\bar1], \hat z_3 = [\bar1\bar1\bar1], \hat z_4 = [\bar111].
	\end{align}
	When $\Bb = 0$, $\ta_a = \phi_a = 0$. The stationary condition under finite $\Bb$ can be obtained by Eq.~\ref{eqHder} up to third order of $B$,
	\begin{align}
		\ta_a = 0,
		\phi_a = -\f{\sqrt6B}{2(4J+J_{DI})} - \f{\sqrt6B^3}{8(4J+J_{DI})^3}.	
	\end{align}
	As $\ta_a = 0$, all spins are confined in $x_ay_a$-planes. Accordingly, the magnetization is $\Mb = \f{\Bb }{4J+J_{DI}}$. 
	
	On the other hand, when $\Bb \parallel [111]$, the symmetry breaking admits the TM. $\Sb_a$ is in Eq.~\ref{eqspin}, whose the local axes are
	\begin{align}
		&\hat x_1 = [525], \hat x_2 = [\bar121], \hat x_3 = [323], \hat x_4 = [111],\n
		&\hat y_1 = [\bar101], \hat y_2 = [101], \hat y_3 = [10\bar1], \hat y_4 = [\bar10\bar1],\n
		&\hat z_1 = [1\bar51], \hat z_2 = [11\bar1], \hat z_3 = [\bar13\bar1], \hat z_4 = [\bar111].
	\end{align}
	Again, $\ta_a,\phi_a=0$ when $\Bb=0$. With finite $\Bb$, from Eq.~\ref{eqHder}, the stationary condition is obtained
	\begin{align}
		&	\ta_1 = \f{16(8J+17J_{DI})}{135(4J+J_{DI})^4}B^3,\n
		&  \ta_2 = \f{4\sqrt2}{3\sqrt3(4J+J_{DI})^2} B^2 - \f{4}{45(4J+J_{DI})^3}B^3,\n
		&	\ta_3 = \f{8(4J-5J_{DI})}{3\sqrt{33}(4J+J_{DI})^4}B^3,\n
		& \ta_4 = \f{-4\sqrt2}{3\sqrt3(4J+J_{DI})^2}B^2 - \f{4}{45(4J+J_{DI})^3}B^3, \n
		& \phi_1 = \f{-4028J+3793J_{DI}}{540\sqrt2 (4J+J_{DI})^4}B^3, \n
		& \phi_2 = \f{2}{3\sqrt3(4J+J_{DI})^2}B^2 -\f{71}{180\sqrt2 (4J+J_{DI})^3}B^3, \n
		& \phi_3 = \f{268J+8707 J_{DI}}{180\sqrt{66} (4J+J_{DI})^4}B^3, \n
		& \phi_4 = \f{-2}{3\sqrt3 (4J+J_{DI})^2}B^2 - \f{71}{180\sqrt2 (4J+J_{DI})^3}B^3.
	\end{align}
	As $\ta_a\neq0$, the spins are away from $x_ay_a$-planes. Accordingly, the TM is
	\begin{align}
		\Mb_{\perp}^{DI} = -\f{4\sqrt2}{3(4+J_{DI}/J)^4}(\f{J_{DI}}{J})(\f{B^3}{J^3}) \hat e_{1\bar21}.
	\end{align}
	
	We analytically calculate the TM by changing $\Bb$ from $[010]$ to $[111]$ and to $[101]$ in sequence (see Fig.~\ref{SFig7}c).  Also, we change $\Bb$ in arbitrary directions shown in Fig.~\ref{SFig7}d.
	$\Mb_{\perp}$ vanishes only at $[010]$ and $[101]$ but is finite otherwise.
	Note that considering the symmetry, the TM vanishes under $[010]$ and $[101]$, but is finite otherwise.
	We plot $\Mb_{\perp}^{DI}$ in units of $J_{DI}/J=0.1$, $B/J =1$ in Figs.~\ref{SFig7}c-d.

	
	\section{Application to experiments \label{Ssec6}}
	Here we apply our theory to the reported experimental results of TM.

	\subsection{CsMnBr$_{3}$}
\begin{figure}[t]
		\centering
		\includegraphics[width=\columnwidth]{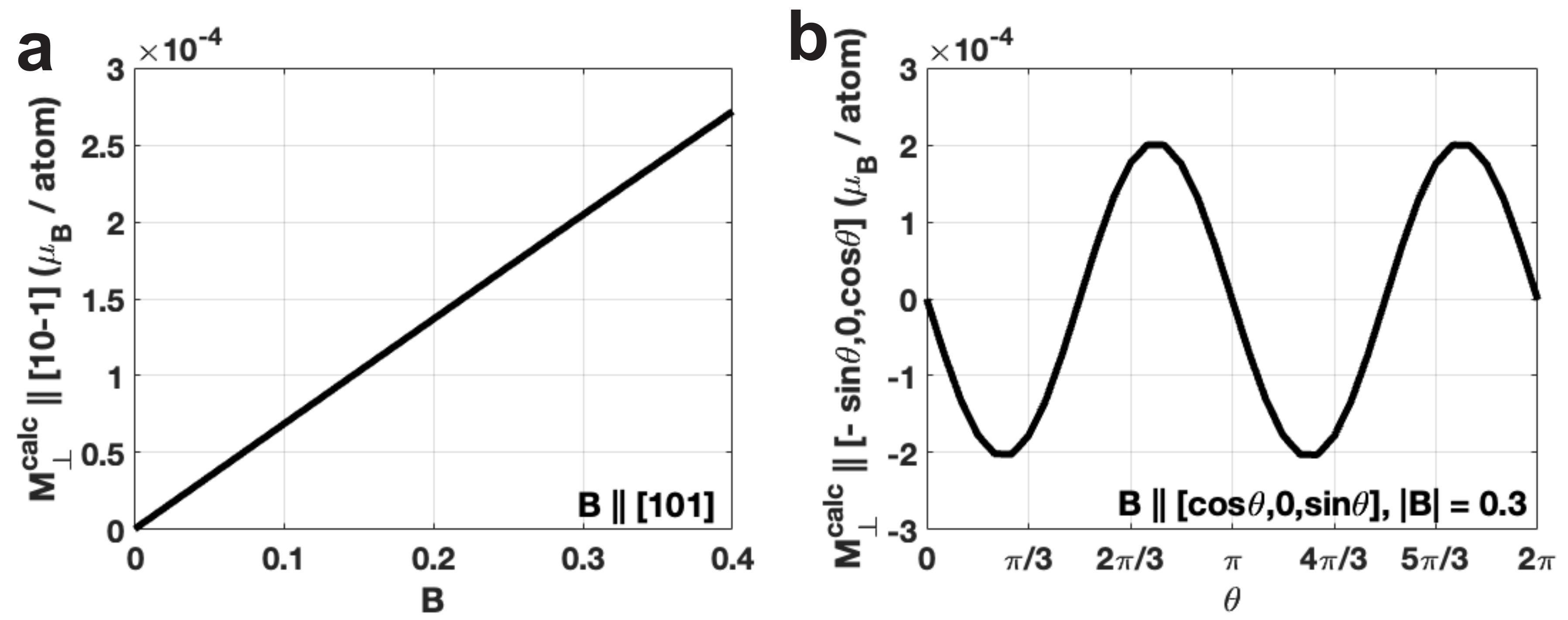}
		\caption{
			Numerical calculations of $\Mb_{\perp}$ using the spin Hamiltonian relevant to CsMnBr$_3$. 
			(a) $\Mb_{\perp}^{calc}$ when $\Bb \parallel [101]$ as a function of $|\Bb|=B$. 
			(b) $\Mb_{\perp}^{calc}$ when $\Bb \parallel [\cos\ta,0,\sin\ta]$ and $B=0.3$, varying $\ta$.
		}
		\label{SFig8}
	\end{figure}

	In CsMnBr$_3$,  $\Mb_{\perp}\propto B$ was observed
	when $\Bb$ is applied within $xz$-plane, 
	unless $\Bb$ is parallel to the $x$ or $z$-axis~\cite{abarzhi1992spin}.
	To numerically calculate $\Mb_{\perp}$, we consider the following spin model relevant to CsMnBr$_3$
	\begin{align}
		H =& J \sum'_{ij} \Sb_i \cdot \Sb_j + J'\sum''_{ij} \Sb_i \cdot \Sb_j \n&+ A \sum_i (S_i^z)^2 - \Bb \cdot \sum_i \Sb_i,\label{eq13}
	\end{align}
	where $J$ ($J'$) is the Heisenberg interaction between intra-layer (inter-layer) nearest neighbors, $A>0$ is the single-ion anisotropy~\cite{chubukov1988quasi}.
	Here $\sum'$ ($\sum''$ ) is the summation over intra-layer (inter-layer) neighbors. The parameters are chosen as $J=2.14, J'=0.005, A =0.0195$.
	The numerically obtained $\Mb_{\perp}^{calc}$ is shown in Figs.~\ref{SFig8}a-b.
	For $\Bb \parallel [101]$, we find $\Mb_{\perp}^{calc} \parallel [10\bar1] \propto B$ (see Fig.~\ref{SFig8}a) while for $\Bb \parallel [\cos\ta,0,\sin\ta]$, we obtain $\Mb_{\perp}^{calc} = 0$ when $\Bb \parallel \hat x$ or $\hat z$ (see Fig.~\ref{SFig8}b),
	which are compatible with the experimental results.
	
	The above numerical results can be understood using the symmetry of CsMnBr$_3$ whose space group is 194 and magnetic space group is 189.225 ($P\bar62'm'$).
	The 6 Mn atoms in the magnetic unit cell form a single spin cluster. The crystalline point group of the unit cell is $D_{3h}$, which is generated by rotations $C_{3z}$, $C_{2x}$, and horizontal mirror $M_z$.

	We have total 18 degrees of freedom of spins in total. The system has 3 cluster dipoles, 7 octupoles, 6 dotriacontapoles, and 2 128-poles. 
	Dipoles are decomposed into $A_2'$ and $E''$, octupoles are decomposed into $A_1'', A_2', A_2'', E'$, and $E''$, dotriacontapoles are decomposed into 2 $E'$ and $E''$, and 128-poles are decomposed into $A_1'$ and $A_1''$.
	The spin configuration of each CMM is shown in SI.
	The ground state is composed of $A_1'$-128-pole, which we denote as $A_1'$-order. (See Fig.~\ref{SFig2}.)
	The magnetic point group of $A_1'$-order is just $D_{3h} (-6m2)$. When $\Bb\parallel z$, the magnetic point group is $\{I,2C_3,M_z,2M_zC_3\}$. 
	On the other hand, when $\Bb \parallel x$, the magnetic point group is $\{I, M_x\}$. Hence, in both cases, $\Mb_{\perp}$ vanishes.
	
	When magnetic field is applied in $xz$-plane, the magnetic point group is $\{I\}$, but the symmetry inverting $\Bb$ is $\{C_{2y}\}$. Using the argument in Sec.~\ref{Ssec1}, the sum of spin changes is
	\begin{align}
		& \Delta S_x(\Bb) = -\Delta S_x(-\Bb), \n & \Delta S_y(\Bb) = \Delta S_y(-\Bb), \n & \Delta S_z(\Bb) = -\Delta S_z(-\Bb),
	\end{align}
	Hence, when $\Bb \parallel [\cos\ta,0,\sin\ta]$, $\Mb_{\perp} \parallel [-\sin\ta,0,\cos\ta]$ is an odd function of $\Bb$. This is consistent with Figs.~\ref{SFig8}a-b.
	
	\subsection{Gd$_2$Ti$_2$O$_7$}
	
	
	Gd$_2$Ti$_2$O$_7$ is a pyrochlore material, where only Gd electrons have magnetism.
	Though the ground state of Gd$^{3+}$ is $^8S_{7/2}$, the strong spin orbit coupling induces nonzero orbital angular momentum and thus a strong single-ion anisotropy appears.
	It has a complicated phase diagram near 1 K; it is known to have a $4$-$k$ structure between 0.75 - 1.05 K~\cite{stewart2004phase}, and have a local-XY structure ($E_2$-order) below 0.75 K~\cite{glazkov2006observation,glazkov2007single}.
	Below 0.75 K, it is reported that $\Mb_{\perp} = 0$ for $\Bb \parallel [001],[110]$ while $\Mb_{\perp} \neq 0$ for $\Bb \parallel [111],[112]$~\cite{glazkov2005single}.
	The reported $\Mb_{\perp}$ is consistent with our numerical calculation in Fig.~\ref{SFig4}.
	That is, at weak field, $\Mb_{\perp} = 0$, but at strong field $\Mb_{\perp}$ becomes nonzero.
	
	First, according to the last row of Table~\ref{TS1}, $\Mb_{\perp}$ vanishes $\Bb \parallel [001]$ or $[110]$, while $\Mb_{\perp}$ appears for $\Bb\parallel [111]$ or $[112]$.
	As the magnetic point group of $E_2$-order is $-42m$,
	when $\Bb\parallel[001]$ $([110])$, the system has $C_{2z}$ $(\ma_{[110]})$, so $\Mb_{\perp} = 0$.
	On the other hand, when $\Bb\parallel[111]$ or $[112]$, all symmetries are broken, so $\Mb_{\perp} \neq 0$.

	\subsection{Eu$_2$Ir$_2$O$_7$}
	
	Eu$_2$Ir$_2$O$_7$ is also a pyrochlore material, where only Ir electrons have magnetism. 
	Because of crystal field and spin-orbital coupling, Ir$^{4+}$ carry the effective spin $J=1/2$.
	The ground state is known to be $A_2$-octupole at low temperature.
	It is reported that when $\Bb$ is applied in $xy$-plane the OM $\Mb_O$ arises~\cite{liang2017orthogonal}.
	$\Mb_O \propto B^2\sin\ta\cos\ta \hat z$ is observed when the field-cooling direction is parallel to $\hat y$.
	
    The OM can be compared with TM.	
    When $\Bb \parallel \hat x$ or $\hat y$, $\Mb_\perp = 0$, according to Table~\ref{TS1} because of $C_2$ symmetry.
	Moreover, considering $S_4 T$ symmetry along $z$-direction, $\Mb_\perp \propto B^2 \sin(2\ta) \hat z$ which is consistent with $\Mb_O$ result.
	We note that in the case of $A_2$-octupole, $\Mb_{\perp}$ and $\Mb_{O}$ show the same $B^2$ and angular dependence. Thus, we cannot rule out the spin canting contribution to OM in the measured data.
	
	\section{The phenomenological model for anomalous and planar Hall Effect\label{Ssec7}}
	
	The physical situation of planar Hall Effect is given in Fig. 4a of our manuscript. When we let $\hat x = [1\bar10], \hat y =[11\bar2], \hat z=[111]$, the electric field is applied along $\hat x$, and the magnetic field is rotating within $xy$-plane ($\Bb=B(\cos\ta,\sin\ta,0)$).
	
	We can acquire a general form of TM by using symmetry analysis. We divide the component of TM into two, $\Mb_{\perp} = \Mb_{\perp,in} + \Mb_{\perp,out}$.
	Along $[111]$, $C_3$ rotation exists. Hence, both in-plane and out-of-plane components obey $a_0 + a_1 \cos 3\ta + a_2 \sin 3\ta$.  During the rotation of magnetic field, the antiunitary mirror is present when $\ta = \pi/6 + n\pi/3$. The antiunitary is spanned by $\hat z$ and $\Bb$, so that the TM can only arise along $\hat z$. Thus, the antiunitary mirror makes the in-plane TM vanishes. This gives the condition of TM components.
	\begin{align}
		&\Mb_{\perp,out} = (A_0 + A_1 \cos 3\ta + A_2 \sin 3\ta)\hat z,\n
		&\Mb_{\perp,in} = (B_1\cos3\ta)\hat p.
	\end{align}
	where $\hat p = (-\sin\ta,\cos\ta,0)$ is the unit vector perpendicular to $\Bb$. 
	The anomalous Hall conductivity is proportional to $\Mb_{\perp,out}$, so
	\begin{align}
		\ma_{xy}^{AHE} \propto (A_0 + A_1 \cos 3\ta + A_2 \sin 3\ta).
	\end{align}
	
	For planar Hall Effect, we first address the Onsager's reciprocal relation. The Onsager's reciprocal relations state that the phenomenological tensors of a certain flow and force in a system out of equilibrium are symmetric. For example, the electrical conductivity under magnetic field $\vec H$ and magnetization $\vec M$ is given by
	\begin{align}
		\ma_{ij}(\V H,\V M) = \ma_{ji}(-\V H, -\V M)
	\end{align}
	Upon this, we assume that the system has a cubic symmetry. By using these two constraints, the current density can be expanded up to the first order of electric field and the second order of magnetic field and magnetization\cite{wang2020antisymmetric,seitz1950note,pippard1989magnetoresistance}. That is,
	\begin{align}
		\V J =& \ma_0 \V E + \ma_1 \V E\times \V H + \ma_2 H^2 \V E + \ma_3 (\V E\cdot \V H)\V H + \ma_4 M^2 \V E \n & + \ma_5 (\V E\times \V M)  + \ma_6 (\V E\cdot \V M)\V M + \ma_7(\V M \cdot \V H)\V E \n & + \ma_8(\V M\times (\V H\times \V E)+\V H\times(\V M \times \V E)) 
	\end{align}
	and the conductivity is
	\begin{align}
		\ma_{ij} =& (\ma_0 + \ma_2 H^2 + \ma_4 M^2 + (\ma_7 - 2\ma_8) (\V M\cdot \V H))\dt_{ij} \n
		& + \ma_1 \ep_{ijk}H_k  + \ma_5 \ep_{ijk} M_k  \n
		& + \ma_3 H_iH_j + \ma_6 M_i M_j + \ma_8(M_iH_j+H_iM_j).
	\end{align}
	The first line indicates the magnetoconductivity. The second line indicates the conventional and anomalous Hall conductivities. The last line gives the phenomenological form of PHC, 
	\begin{align}
		\ma_{xy}^{PHE} =& \ma_3 H_xH_y + \ma_6 M_x M_y \n&+ \ma_8(M_xH_y+H_xM_y). \label{eq4}
	\end{align}
	This is the equation that we are based on. Let $\Mb=\Mb_{\perp,in}$ above, then the angular dependence of PHC is
	\begin{align}
		\ma_{xy}^{PHE} =& A_1 \cos \ta + A_2 \cos 5\ta \n&+ B_1 \sin2\ta + B_2 \sin 4\ta + B_3 \sin 8\ta.
	\end{align}

%


\end{document}